\documentclass[aps,prb,twocolumn,superscriptaddress,showpacs]{revtex4-1}

\usepackage{graphicx}
\usepackage{color,soul}
\usepackage{latexsym}
\usepackage{subfigure}

\begin{document}

\title{Temperature and field evolution of site-dependent magnetism in $\epsilon$-Fe$_2$O$_3$ nanoparticles}

\author{Richard Jones}
\affiliation{Department of Physics and Astronomy, University of Manitoba, Winnipeg, Manitoba, R3T 2N2, Canada}

\author{Rachel Nickel}
\email{nickelr@myumanitoba.ca}
\affiliation{Department of Physics and Astronomy, University of Manitoba, Winnipeg, Manitoba, R3T 2N2, Canada}

\author{Palash K. Manna}
\affiliation{Department of Physics and Astronomy, University of Manitoba, Winnipeg, Manitoba, R3T 2N2, Canada}
\affiliation{School of Physical Sciences, NISER Bhubaneswar, Odisha 752050, India}

\author{J. Hilman}
\affiliation{Department of Physics and Astronomy, University of Manitoba, Winnipeg, Manitoba, R3T 2N2, Canada}

\author{Johan van Lierop}%
\email{Johan.van.Lierop@umanitoba.ca}
\affiliation{Department of Physics and Astronomy, University of Manitoba, Winnipeg, Manitoba, R3T 2N2, Canada}
\affiliation{Manitoba Institute for Materials, University of Manitoba, Winnipeg, Manitoba, R3T 2N2, Canada}


\begin{abstract}

8~nm $\epsilon$-Fe$_2$O$_3$ nanoparticles exhibit a spin reorientation transition that begins at 150~K which is a hallmark of this unique iron-oxide polymorph.  We find that the change from the high to low temperature magnetic structures has been suppressed by $\sim$50~K.  At the spin reorientation temperature, a change of the field-dependent response of the tetrahedral sites in intermediate field strengths (0.25 -- 1.5~T) indicates that a collective tetrahedral distortion occurs to which the octahedral sites adjust, altering the magnetic anisotropy.  An abrupt step in the hyperfine parameters' temperature dependencies, especially at 125~K for the hyperfine field associated with the Fe$_4$ tetrahedral sites, suggests strongly a change in the superexchange pathways are responsible for the spin reorientation.

\end{abstract}

\maketitle


\section{Introduction}

$\epsilon$-Fe$_2$O$_3$ is the least common and understood iron oxide polymorph.  Initial reports specifying this phase, distinct from Fe$_3$O$_4$, and $\alpha$- and $\gamma$-Fe$_2$O$_3$, appeared in the 1930s\cite{Forestier.1934}.  But, it was not until the late 1990s\cite{Tronc.1998} that a silica matrix surrounding $\gamma$-Fe$_2$O$_3$ parent nanoparticles was identified as the method to permit phase-pure synthesis of this intermediate $\epsilon$ phase between metastable $\gamma$-Fe$_2$O$_3$ and thermodynamically stable $\alpha$-Fe$_2$O$_3$.  $\epsilon$-Fe$_2$O$_3$ has attracted significant interest recently due to its potential of providing large coercive fields \cite{Ohk.2012,Ohkoshi.2013,Gich.2005,Tseng.2009}, a characteristic unique from the other iron oxides.  In addition, $\epsilon$-Fe$_2$O$_3$ is a multiferroic with an appreciable room temperature ferrimagnetism and ferroelectricity\cite{Gich.2014}, making it an interesting complimentary candidate material to BiFeO$_3$ (BFO), the well-known room temperature multiferroic\cite{Guo.2017} that has an antiferromagnetic-like incommensurate cycloidal-type spin configuration that results in no useful magnetization at room temperature, hampering the development of applications\cite{Xu.2018}.

Despite the wealth of research into the iron oxides, a detailed understanding of the $\epsilon$-phase's structure was not achieved until 1998 by Tronc et al.\cite{Tronc.1998}, reflecting the difficulty of phase-pure synthesis. Sol-gel techniques have largely been adopted, allowing for relatively precise control over particle size, a critical factor in achieving thermodynamic stability of the $\epsilon$-phase\cite{Ohk.2005}.  To change the shape and magnetocrysalline anisotropy contributions to the magnetism and thereby increase the coercivity, several groups have focused on site-substitution with metal ions such as aluminum, gallium and indium\cite{Ohk.2005, Ohk.2012, Ohkoshi.2013, Tseng.2009, Gich.2006, Tucek.2011}. Such systems show coercive fields very often exceeding 2~T at room temperature, due in part to rod-like crystallization encouraged by the preferential adsorption at the impurity sites, as well as larger spin-orbit coupling enabled with the heavy metal ions. There has also been recent success in the direct conversion from magnetite (Fe$_3$O$_4$) to $\epsilon$-Fe$_2$O$_3$ via a nano-size wet process\cite{Ohk.2017}, though the conversion largely results in maghemite ($\gamma$-Fe$_2$O$_3$). 

 $\epsilon$-Fe$_2$O$_3$ crystallizes into an orthorhombic cell structure in the space group $Pna2_1$\cite{Gich.2006}.  The unit cell parameters at room temperature have been reported as $a=5.095$~\AA, $b=8.789$~\AA, and $c=9.437$~\AA\cite{Ohkoshi.2013}, and it is comprised of four unique iron sites. A site labelling convention has emerged that tracks the level of distortion of the sites at 200~K\cite{Tucek.2010}, where two octahedrally coordinated sites, Fe$_1$ and Fe$_2$ show similar, high levels of distortion, while another octahedral, Fe$_3$, and a tetrahedral, Fe$_4$ site present relatively little distortion.

$\epsilon$-Fe$_2$O$_3$ is a ferrimagnet,  although there exists some disagreement in the literature regarding the collinearity of the sublattice magnetizations. A collinear ferrimagnetic description has been identified by many groups\cite{Ohkoshi.2013, Tseng.2009, Tucek.2011}, where a weakly compensating tetrahedral sublattice magnetization results in the small ferromagnetic component along the $a$-axis of the unit cell.  Below a transition temperature of $\sim$125~K, however, the spin configuration turns into an incommensurate structure involving square-wave oscillation of the sublattice magnetizations.  An alternative description of sinusoidally oscillating sublattice magnetizations has also been presented based on neutron powder diffraction measurements\cite{Gich.2006}. Additionally, a spin-canted description\cite{Kurmoo.2005} as the origin of the ferrimagnetism has been inferred from the superexchange pathways associated with the 16 iron sites in the unit cell, where chains of Fe$_1$-Fe$_2$-Fe$_3$ octahedra form ferromagnetic units that are antiferromagnetically coupled to the Fe$_4$ tetrahedra. The tetrahedral sites experience spin frustration due to Fe$_4$-Fe$_4$ antiferromagnetic superexchange, resulting in a canted ground state. In this description, the transition at 125~K is associated with a change in the canting angle\cite{Kurmoo.2005}.  The transition is most notably marked by a near complete collapse of the coercive field around 75--80~K, followed by a reemergence around 125~K, indicating a change in the nature of the magnetic anisotropy and associated spin and electronic configurations of the Fe sites.

We have examined how the transition described above impacts $\sim$8~nm $\epsilon$-Fe$_2$O$_3$ nanoparticles that are approximately half the size of previously studied systems which are phase pure and reasonably monodisperse (synthesis of this material is an ongoing challenge).  These smaller nanoparticles provide an ideal test-bed for examining the interplay between the nanomagnetism of finite size effects (e.g. surface spin disorder) and a magnetic spin transition that is unique amongst the iron oxides.  Considering the unit cell of $\epsilon$-Fe$_2$O$_3$ is smaller compared to the other ferrimagnetic iron oxides ($\sim$0.42~nm$^3$ vs $\sim$0.58~nm$^3$ for $\gamma$-Fe$_2$O$_3$ and Fe$_3$O$_4$) and its symmetry is low such that spin interactions are quite complex\cite{Xu.2018}, $\epsilon$-Fe$_2$O$_3$ is likely even more susceptible to finite-size effects on its magnetism\cite{finitesize}.  Indeed, we find that due to the approximately three-fold increase in surface-to-volume Fe$^{3+}$ (compared to the larger, e.g. 25~nm, crystallites studied) that the transition from the high to low temperature magnetic structures had been suppressed to 40~K in comparison to the 90~K observed in larger (e.g.~25~nm) $\epsilon$-Fe$_2$O$_3$ nanoparticles.  The effect of surface spin disorder is revealed by an exchange bias field whose temperature dependence is linked to the transition.  Site-specific magnetism identified by x-ray magnetic circular dichroism and M\"ossbauer spectroscopy show that the temperature dependence of the orbital and spin magnetism maps onto the observed changes in overall magnetic anisotropy. Also, an abrupt step in the hyperfine fields associated with the Fe$_4$ tetrahedral sites suggests strongly a change in that superexchange pathway through the oxygen ions is responsible.  The definite and unique thermal evolution of the Fe-sites' magnetism along with an apparent lack of inter-site disorder suggests a collective transition of electronic localization properties.


\section{Experimental Methods}

The nanoparticles were synthesized following the protocol developed by Ohkoshi et al.\cite{Ohkoshi.2013}, involving a colloid of reverse micelles impregnated with iron nitrate nonahydrate (Alfa Aesar), and silica coated using a sol-gel approach\cite{Hench.1990}.  We did not attempt to promote rod growth via preferential adsorption of heavy metal ions as in previous works, and instead modified the density of the alkane dispersive media by using n-decane (Fisher Chemical) instead of n-octane (n-hexane and n-dodecane were also used, but resulted in a small $\alpha$-Fe$_2$O$_3$ component, consistent with the sol-gel stage forming micelles too large for $\epsilon$-Fe$_2$O$_3$ thermodynamic stability, where the condition for a given phase is related to the particle diameter\cite{Ohkoshi.2013}).  In brief, a reverse micelle solution was formed from 18.3~ml of n-decane mixed with 6~ml of DI water, 3.7~mL of 1-Butanol (Fisher), 4.05~g Cetrimonium bromide (CTAB; Fisher), and 0.9~g Fe(NO$_3$)$_3$-9H$_2$O (Alfa Aesar).  A second reverse micelle solution of made with the same amounts of decane, DI water and 1-Butanol with 2~ml of aqueous ammonia (20-30\%; Aldrich) and 3.7~g CTAB.  This second reverse micelle solution was added drop-wise to the first one under vigorous stirring over 10 minutes, and stirred for an addition 30 minutes.  Then, a gel of 1.0~ml of tetraethyloxysilane (TEOS; Fisher) was added to the sol of dispersed reverse micelle solutions, and mixed for 18~hours.  The solution was then centrifuged, and the precipitate washed in equal parts ethanol and chloroform to remove excess organics and surfactants, and air dried at 70$^{\circ}$C for 12~hours.  The resulting transformation product was vacuum dried for two hours and sintered in a tube furnace at 1000$^{\circ}$C for four hours.  For imaging purposes, the majority of the silica matrix was dissolved by stirring the $\epsilon$-Fe$_2$O$_3$ nanoparticles in a 2M NaOH bath for 24 hours.

X-ray diffraction patterns were collected using a Bruker D8 Discover with DaVinci using Cu K$_{\alpha}$ radiation with a Bragg-Brentano geometry.  For patterns collected under ambient conditions, a rotating stage was used with the sample mounted onto a zero-background quartz slide.  A G\"obel mirror was used on the primary optics, knife edges were used to minimize air-scattering, and a motorized slit and a Ni filter was used in front of the secondary detector Lynxeye optics in 1D mode.  The low temperature x-ray diffraction patterns were collected using a Phenix closed-cycle refrigeration stage. 

Transmission electron microscopy (TEM) imaging was performed using a FEI Talos F200X S/TEM.  The nanoparticles were prepared by dropping a mixture of nanoparticle solution diluted in hexanes onto a copper coated carbon grid. 

X-ray absorption spectroscopy (XAS) and x-ray magnetic circular dichroism (XMCD) measurements were done at beam line 4-ID-C of the Advanced Photon Source in a liquid helium cryostat with powder samples mounted on carbon tape onto a cold finger in a 7~T (maximum) field magnet. All XMCD spectra were collected in total electron yield mode and are XAS normalized as well as averaged over both field polarity measurements to remove non-magnetic artifacts.

Transmission M\"ossbauer spectra were collected between 10 and 300~K using a Janis SHI-850 closed cycle refrigeration system and a WissEl constant acceleration spectrometer with a 10-GBq $^{57}$Co$\bf{Rh}$ source. The source drive velocity was calibrated using a 6~$\mu$m thick $\alpha$-Fe foil at room temperature.

Magnetometry and susceptometry experiments were performed using a Quantum Design magnetic properties measurement system (MPMS XL-5). Hysteresis loops were measured from 2~K to 400~K from $\pm50$~kOe after cooling in 50~kOe field, or a 0~Oe field (having quenched the magnet at 300~K beforehand, and using the field-profiling and cancelling capability).  Zero-field cooled (ZFC)/field cooled (FC) DC susceptibility measurements were done by cooling from 400 to 2 K and subsequent measurements using 100 and 1000~Oe applied fields, and AC susceptibility measurements were done using a 2.5~Oe oscillating applied field from 10~Hz to 1k~Hz from 2~K to 300~K.  The nanoparticle sample was mounted in a Supracil quartz NMR tube.  Energy dispersive x-ray (EDX, collected with a FEI Nova NanoSEM 450) data of the nanoparticles were used to estimate the Fe-oxide-to-SiO$_2$ mass fractions so as to normalize the magnetometry and susceptometry measurements..


\section{Results and Discussion}

\begin{figure}[t!]

\includegraphics[scale=0.375]{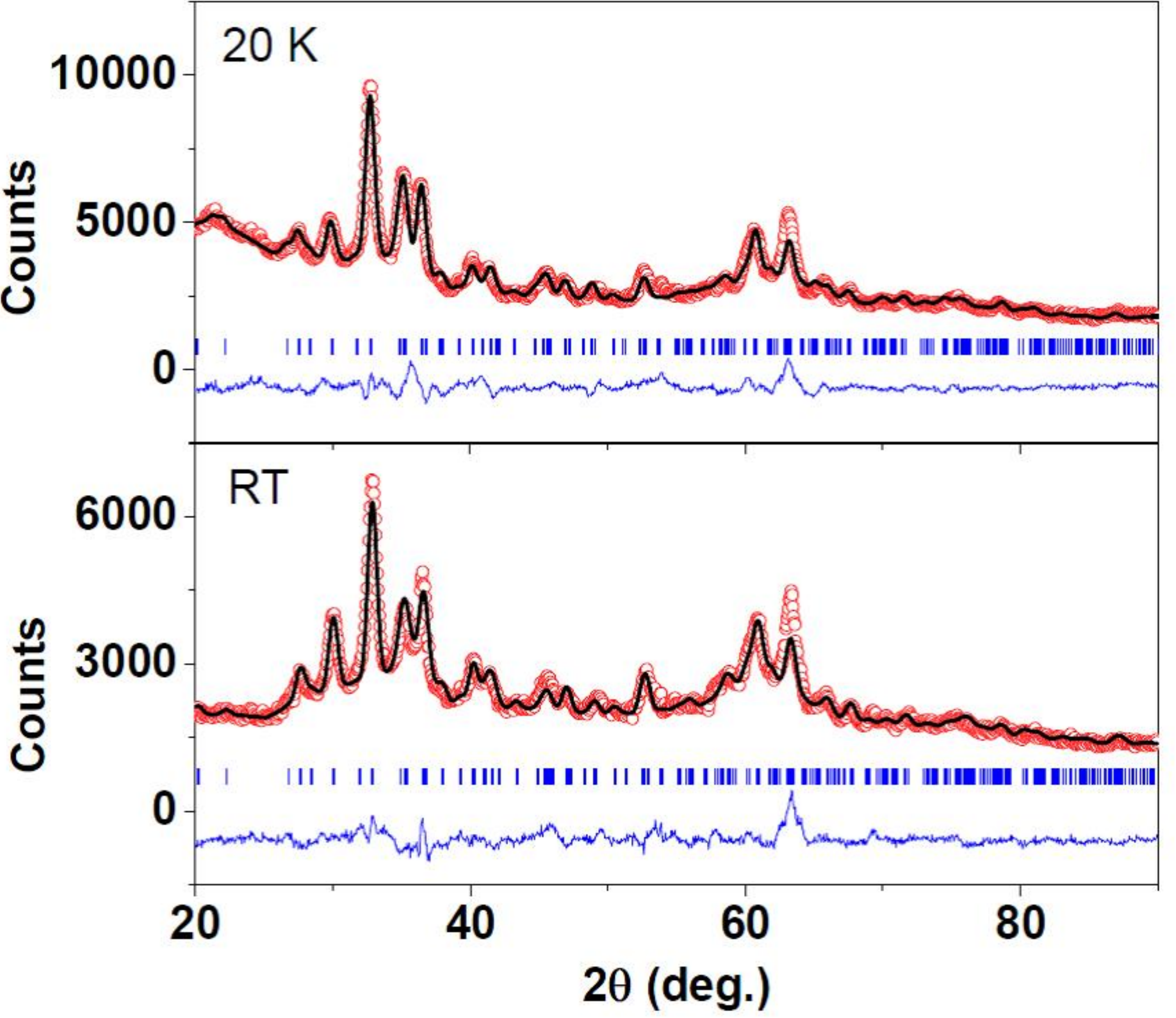}

\caption{(colour online)  XRD pattern (red $\circ$) at 20~K (top) and room temperature (bottom) with refinements (solid black line), as described in the text.  Residuals and Bragg markers are in blue.\label{fig:xrd}}

\end{figure}

\begin{figure}[b!]

\subfigure[\ ]{\includegraphics[scale=0.45]{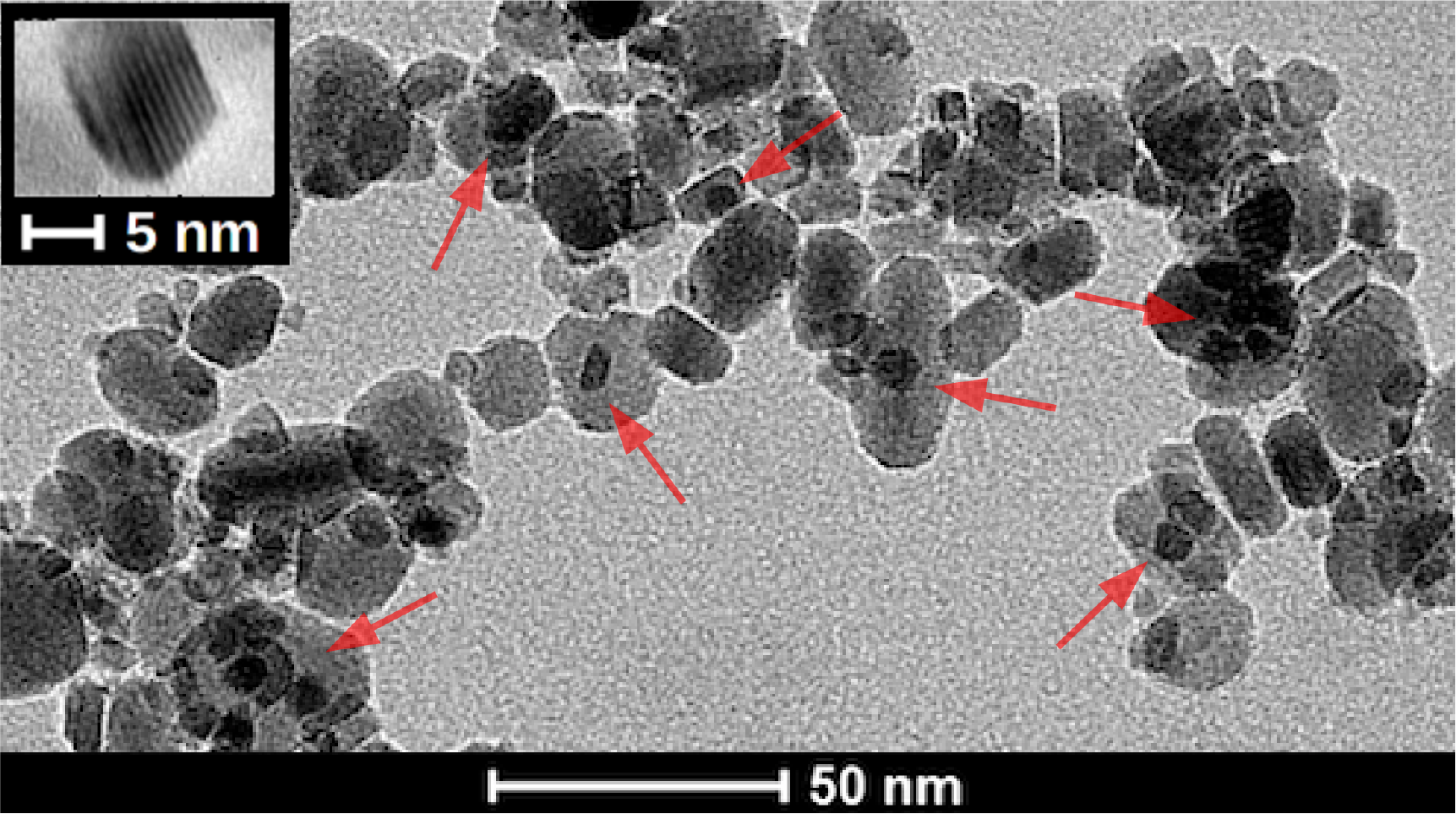}}

\subfigure[\ ]{\includegraphics[scale=0.275]{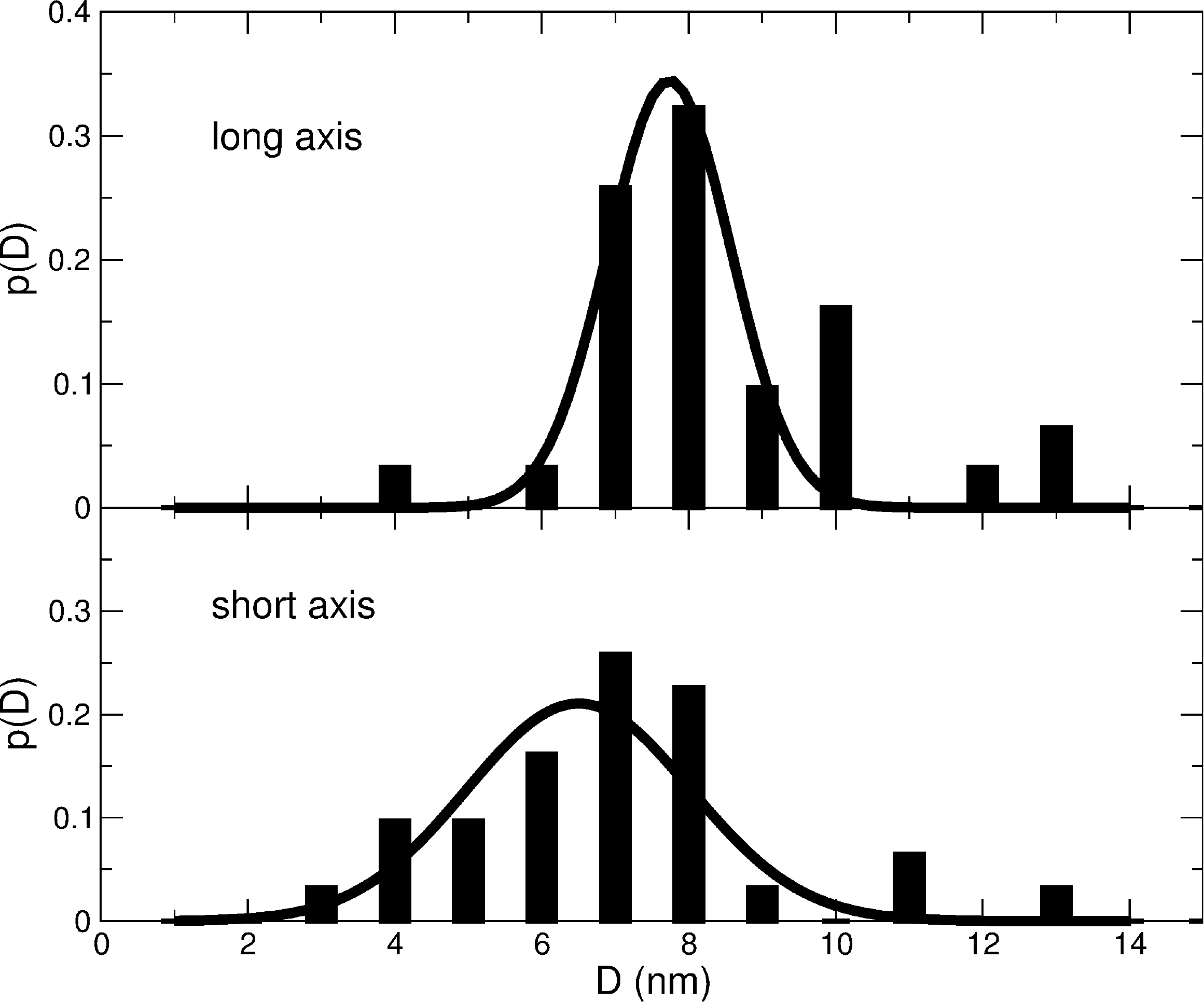}}

\caption{a) A representative TEM image of the $\epsilon$-Fe$_2$O$_3$ nanoparticles where the majority of the silica coating was stripped (post synthesis, as described in the text) to permit imaging.  Representative nanoparticles in the remaining SiO$_2$ shells are identified with red arrows.  The inset shows a single nanoparticle with the lattice planes.  b)  Distributions of the long- and short-axis of $\sim$200~nanoparticles (solid lines are a guide to the eye).\label{fig:TEM}} 

\end{figure}

Representative XRD patterns collected between 20~K and room temperature of the $\epsilon$-Fe$_2$O$_3$ nanoparticles with the majority of the SiO$_2$ removed are shown in Fig.~\ref{fig:xrd}.  The contribution from the remaining amorphous SiO$_2$ was included in the patterns' backgrounds during the Reitveld refinement process.  Using the FullProf package\cite{FullProf}, refinements were done using the $Pna2_1$ space group with site occupancies set to the values of $\epsilon$-Fe$_2$O$_3$\cite{Gich.2006} (i.e. equivalent occupation in the unit cell of each of the four iron sites) and the fitted lattice parameters were $a$=5.093$\pm$0.002~\AA, $b$=8.874$\pm$0.004~\AA~and $c$=9.500$\pm$0.003~\AA.  The  Scherrer broadening incorporated into the refinements identified a volume averaged crystallite size of $\sim$8~nm.  Unlike previous results on larger $\sim$25~nm crystallites\cite{Gich.2006, Tseng.2009}, there was no observable difference between the ambient temperature pattern and those collected at 50~K increments down to 20~K\cite{XRDcomment,Sakurai.2005}.  Previous reports on larger $\epsilon$-Fe$_2$O$_3$ crystallites (e.g. see Tseng et al.\cite{Tseng.2009} and references therein) identified an observable temperature dependence of $a/c$ and $b/c$ changing by 0.03--0.04\% between 200 and 10~K.  The XRD patterns of the 8~nm nanoparticles are Scherrer broadened to an extent that likely obscures such small changes in the lattice with cooling.  Figure~\ref{fig:TEM}a shows a representative TEM image of the $\epsilon$-Fe$_2$O$_3$ nanoparticles with most of the SiO$_2$ matrix removed.  The particles are mostly parallelepiped shaped, with a tendency towards cubes.  Due to the range of shapes, we present the distribution of long- and short-axis `sides' from TEM images (Fig.~\ref{fig:TEM}b) determined from ImageJ\cite{ImageJ} analysis.  Nanoparticles appear to be well crystallized with clear images of the lattice planes (example in Fig.~\ref{fig:TEM}a inset). 

\begin{figure}[t!]

\includegraphics[scale=0.33]{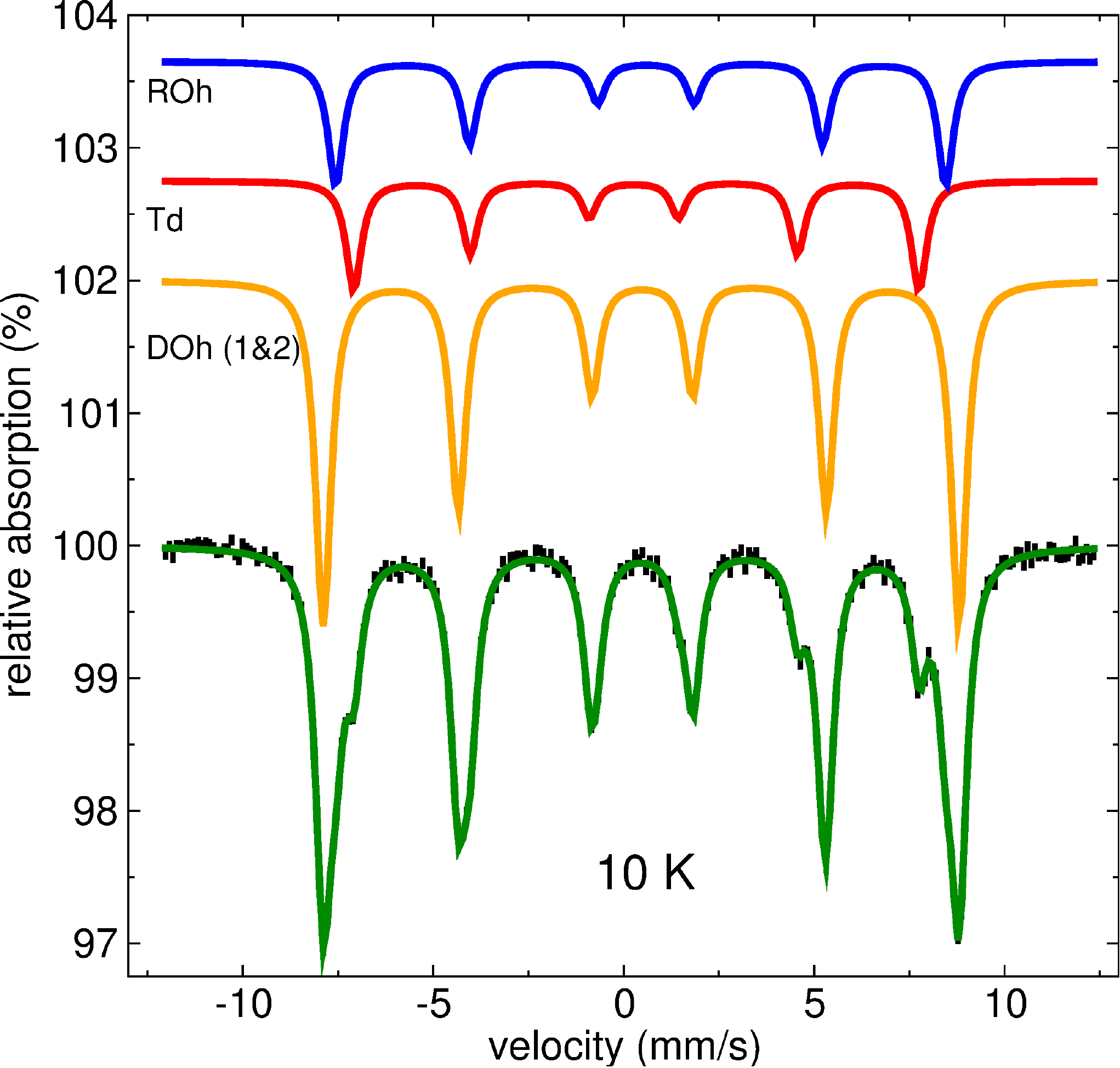}

\caption{(colour online) 10~K transmission M\"ossbauer spectrum with the fit (solid green line) and spectral components presented above the data.  Fe$_{\rm RO}$ (blue) and Fe$_{\rm DO(1\&2)}$ (orange) denote the regular octahedral (Fe$_3$) and distored octahedral (Fe$_1$ and Fe$_2$) sites, and Fe$_{\rm T}$ (red) is the tetrahedral Fe$_4$ site component. \label{fig:Mossy_10K}}
\end{figure}

The 10~K M\"ossbauer spectrum (Fig.~\ref{fig:Mossy_10K}) presents hyperfine parameters that agree with previous measurements\cite{Gich.2006}.  The spectrum is well described by three subspectra with Lorentzian FWHM $\Gamma$=0.247$\pm$0.008~mm/s (about twice the source's natural $\Gamma$=0.133$\pm$0.002~mm/s) that reflects the intrinsic chemical and structural disorder due to finite size effects in the 8~nm nanoparticles.  The disordered octahedral sites (Fe$_1$ and Fe$_2$) make up 55$\pm$3\% of the absorption, and have a hyperfine field $B_{hf}$=51.83$\pm$0.06~T and isomer shift, $\delta$=0.814$\pm$0.006~mm/s.  The broadened $\Gamma$ made resolving the individual Fe$_1$ and Fe$_2$ sites problematic (fits would have both sites settle into the same hyperfine parameters, although with increasing temperature the sites' subspectra were observable as discussed below).  The regular, ordered octahedral site (Fe$_3$) makes up 22$\pm$4\% of the relative absorption with a $B_{hf}$=49.73$\pm$0.08~T and $\delta$=0.856$\pm$0.006~mm/s, and experiences local distorted coordination (as expected from the crystal structure\cite{Gich.2006}) that provides an electric field described by a quadrupole shift $\Delta$=-0.14$\pm$0.03~mm/s.  The tetrahedral Fe$_4$ site makes up the rest of the spectrum with a $B_{hf}$=46.08$\pm$0.02~T and $\delta$=0.640$\pm$0.002~mm/s.

\begin{figure}[t!]

\includegraphics[scale=0.33]{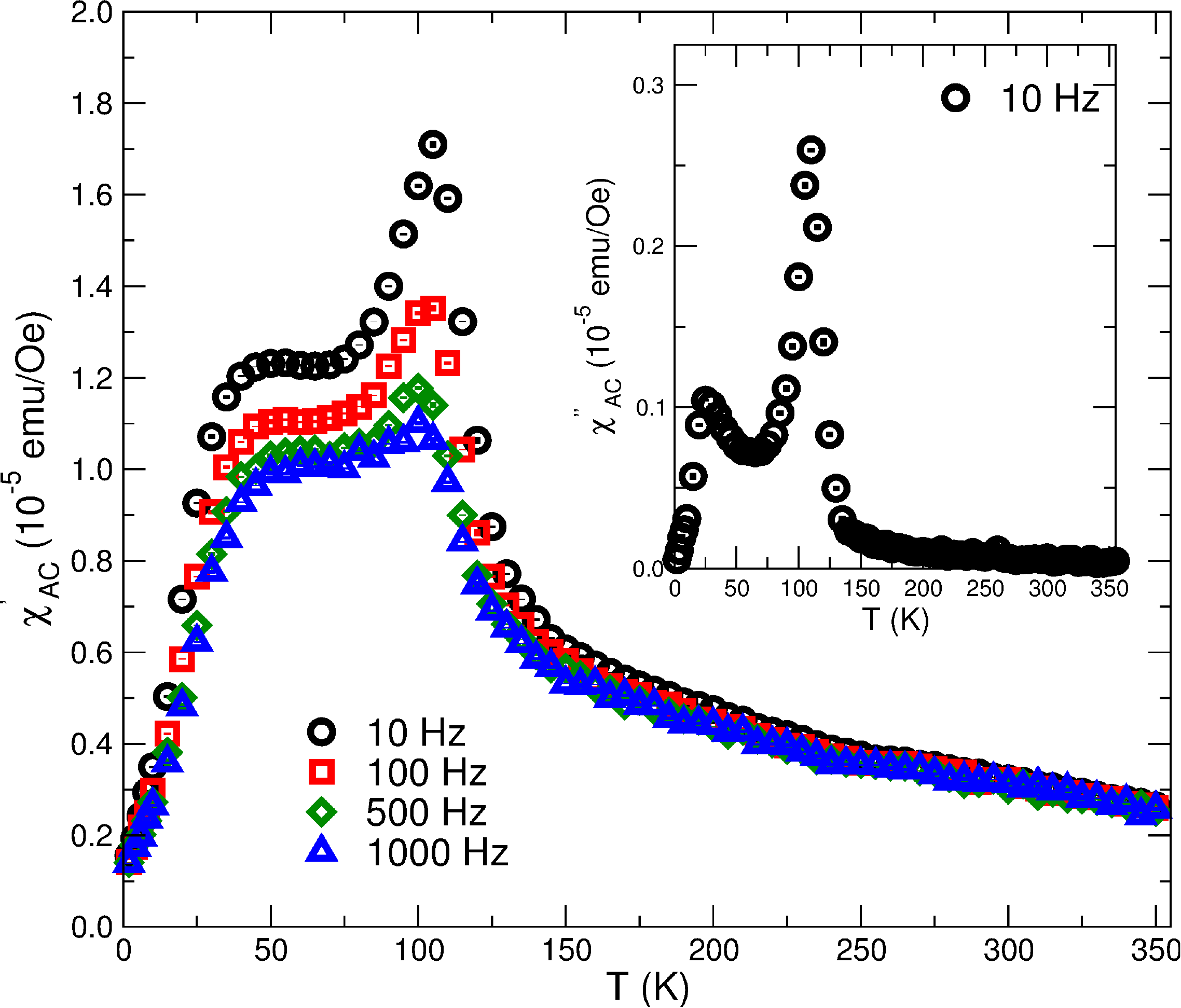}

\caption{(colour online) Temperature and frequency (10~Hz $\circ$, 100~Hz $\Box$, 500~Hz $\Diamond$  and 1000~Hz $\triangle$) dependence of the in-phase $\chi'(T,\nu)$ and out-of-phase ($\chi''(T)$ at 10~Hz - inset)  AC susceptibility of the $\epsilon$-Fe$_2$O$_3$ nanoparticles.\label{fig:Xac}}

\end{figure}

\begin{figure}[b!]

\includegraphics[width=0.45\textwidth]{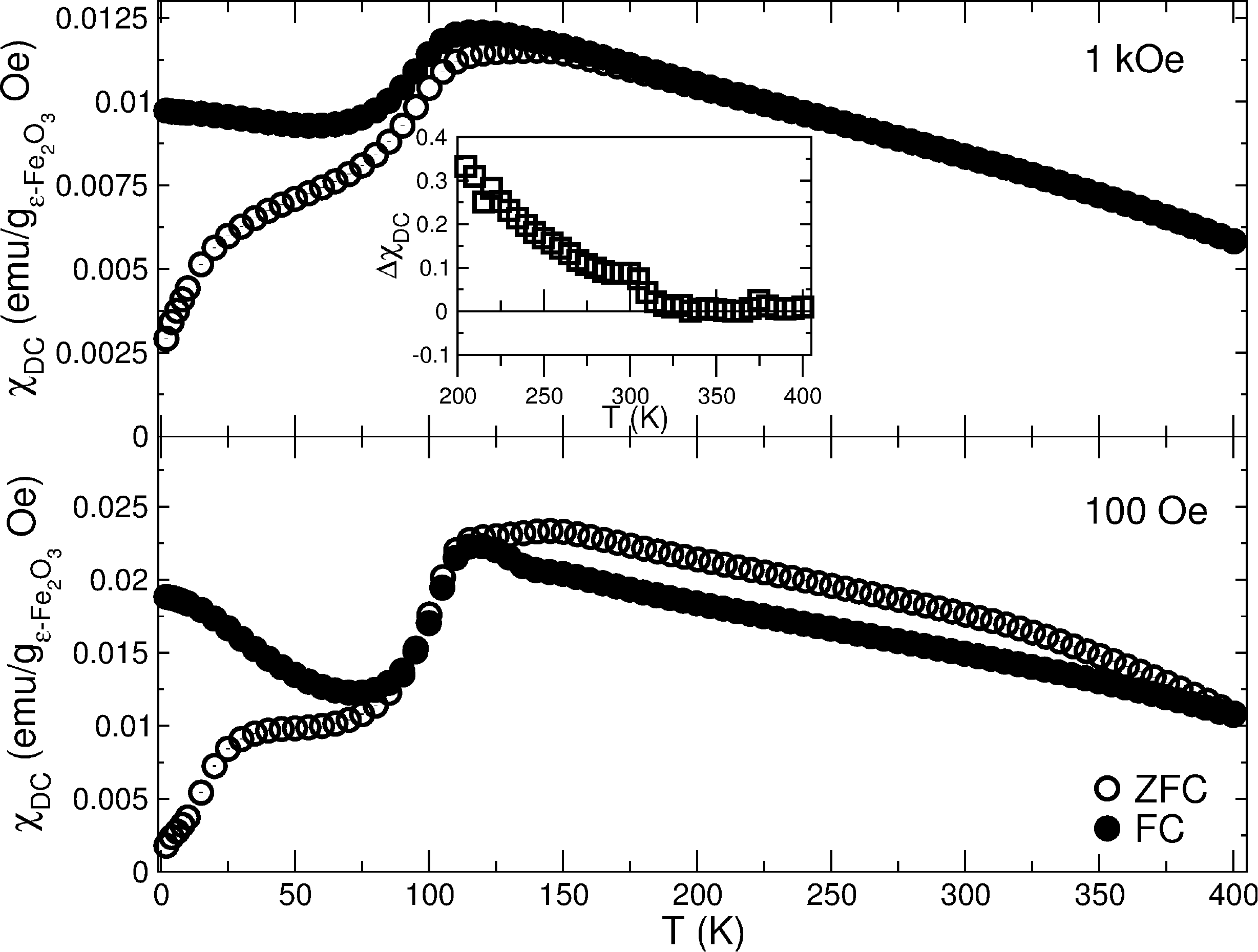}

\caption{Temperature dependence of the zero field-cooled (ZFC) and field-cooled (FC) DC susceptibility $\chi_{\rm DC}$ in 100~Oe (bottom) and 1~kOe (top) fields.  The inset shows the difference between 1~kOe field-cooled and zero field-cooled susceptibilities, $\Delta \chi_{\rm  DC}$ at temperatures between 200 and 400~K.\label{fig:chiDC}}

\end{figure}

\begin{figure*}

\subfigure[\ ]{\includegraphics[scale=0.175]{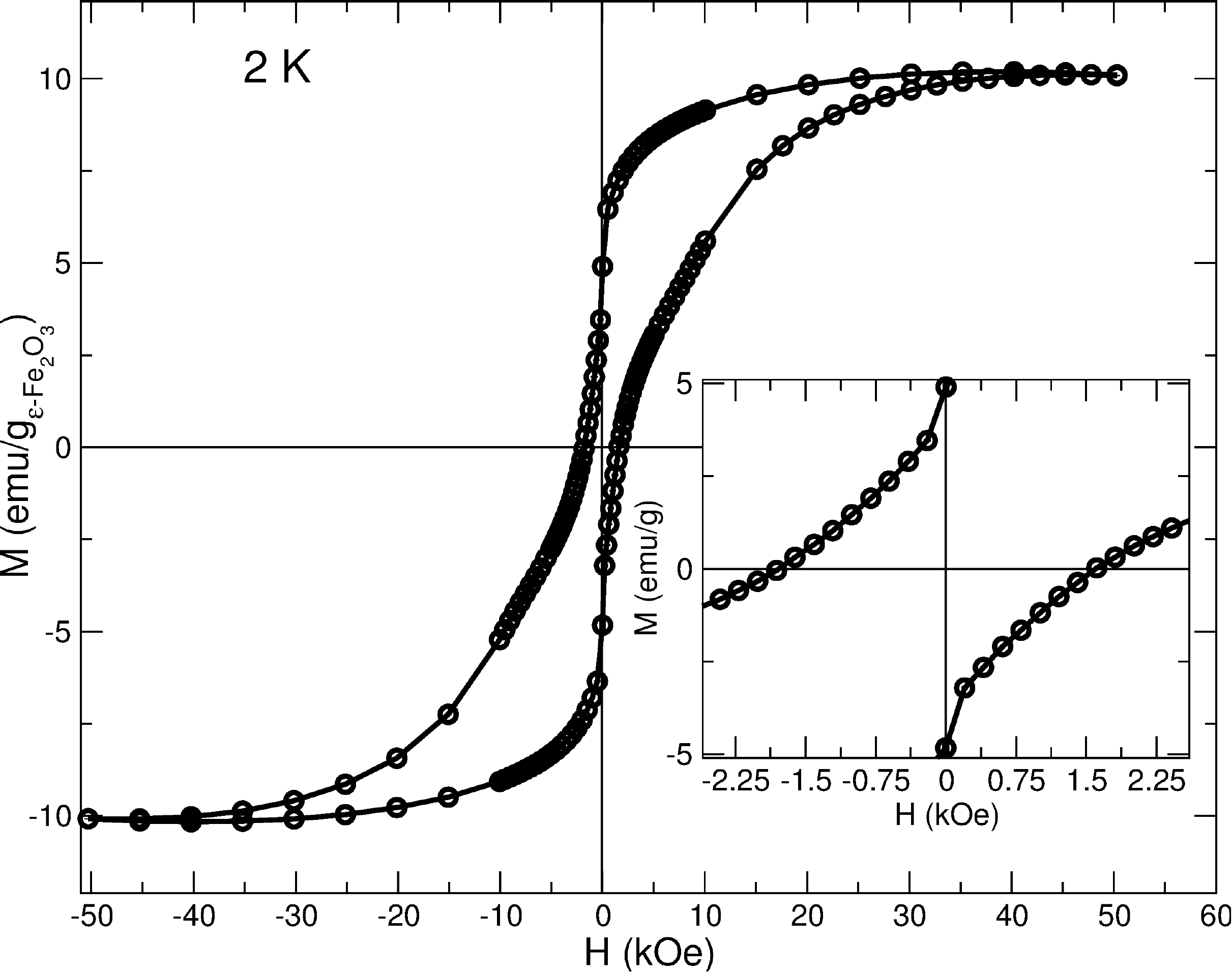}}
\subfigure[\ ]{\includegraphics[scale=0.175]{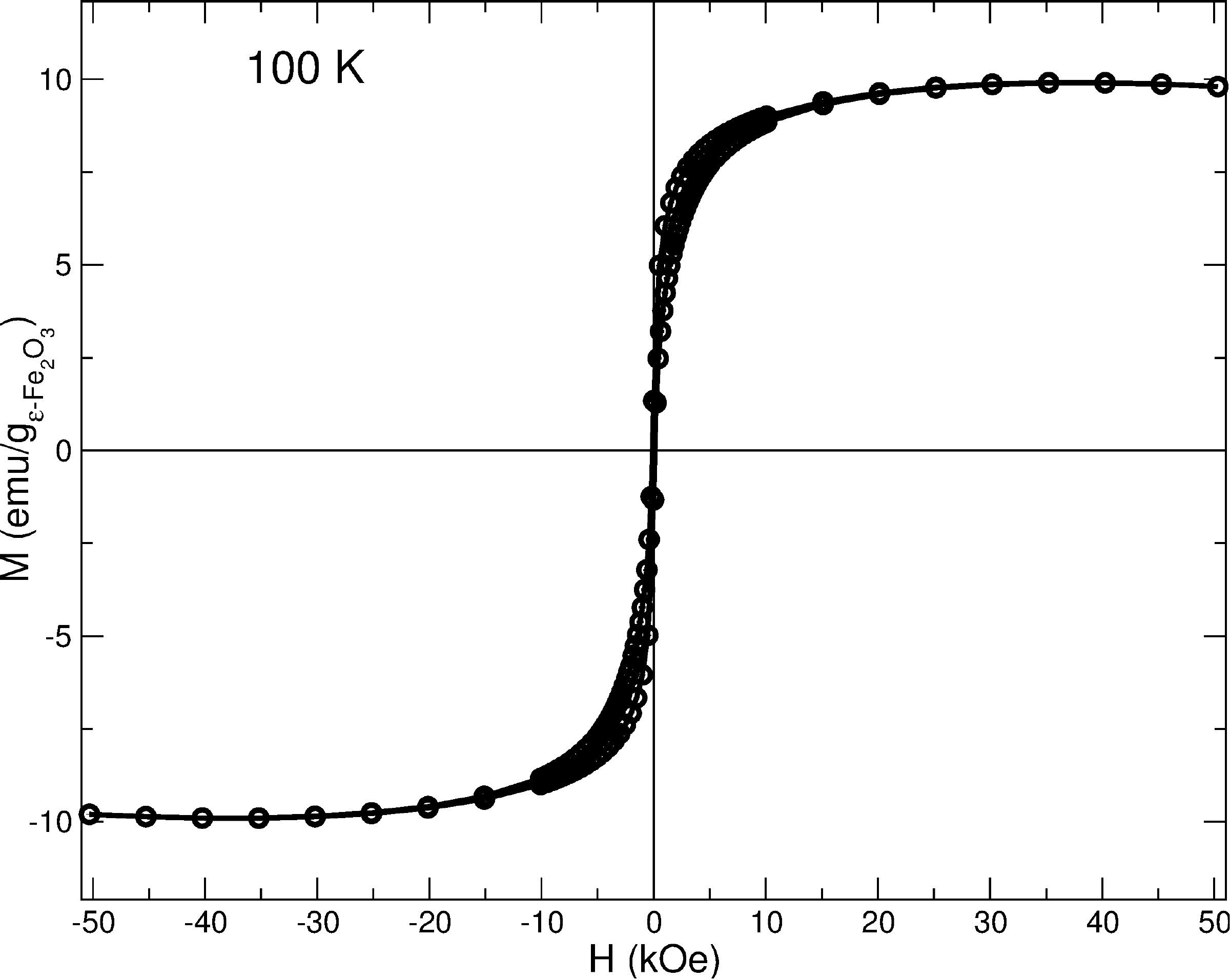}}
\subfigure[\ ]{\includegraphics[scale=0.175]{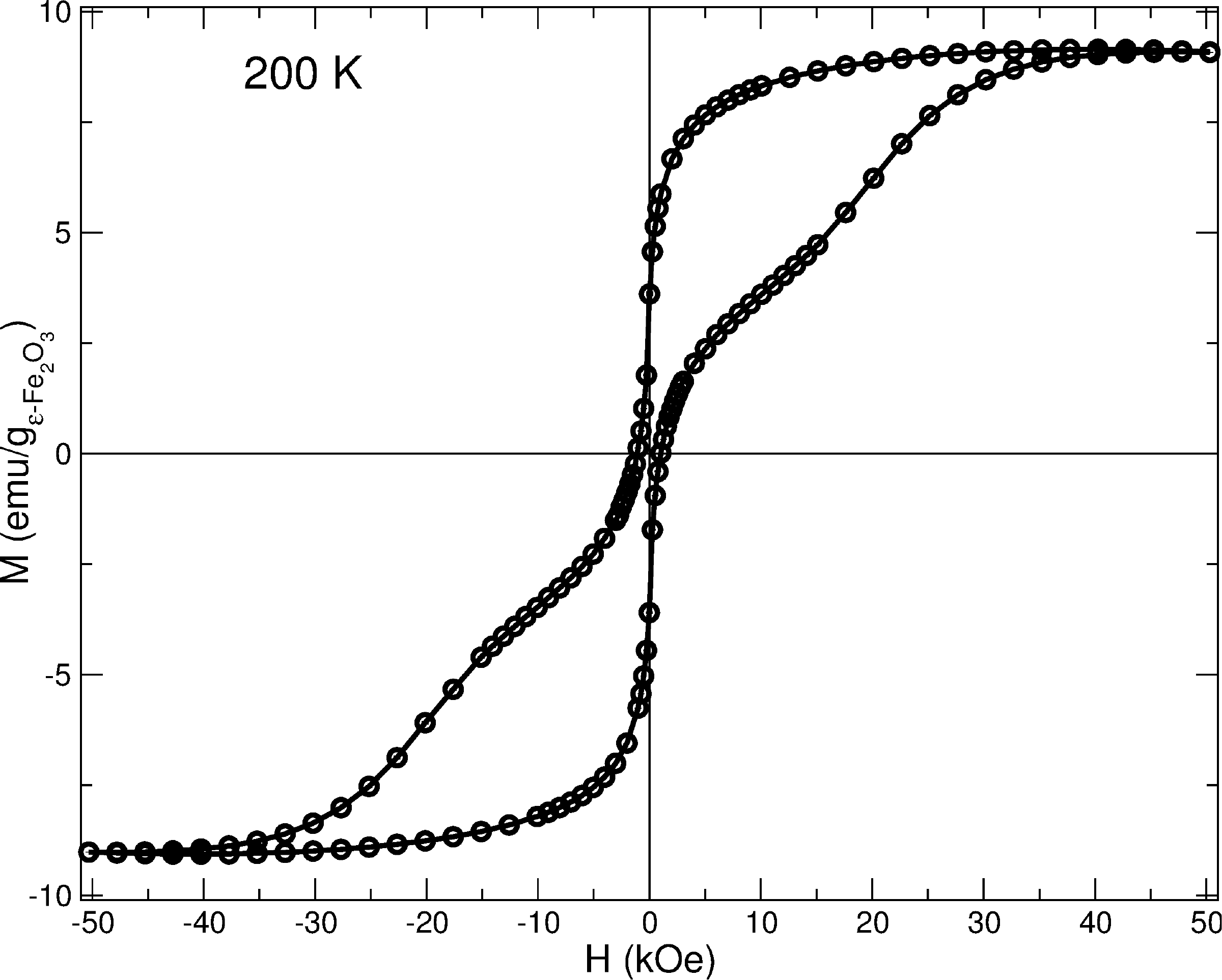}}
\subfigure[\ ]{\includegraphics[scale=0.175]{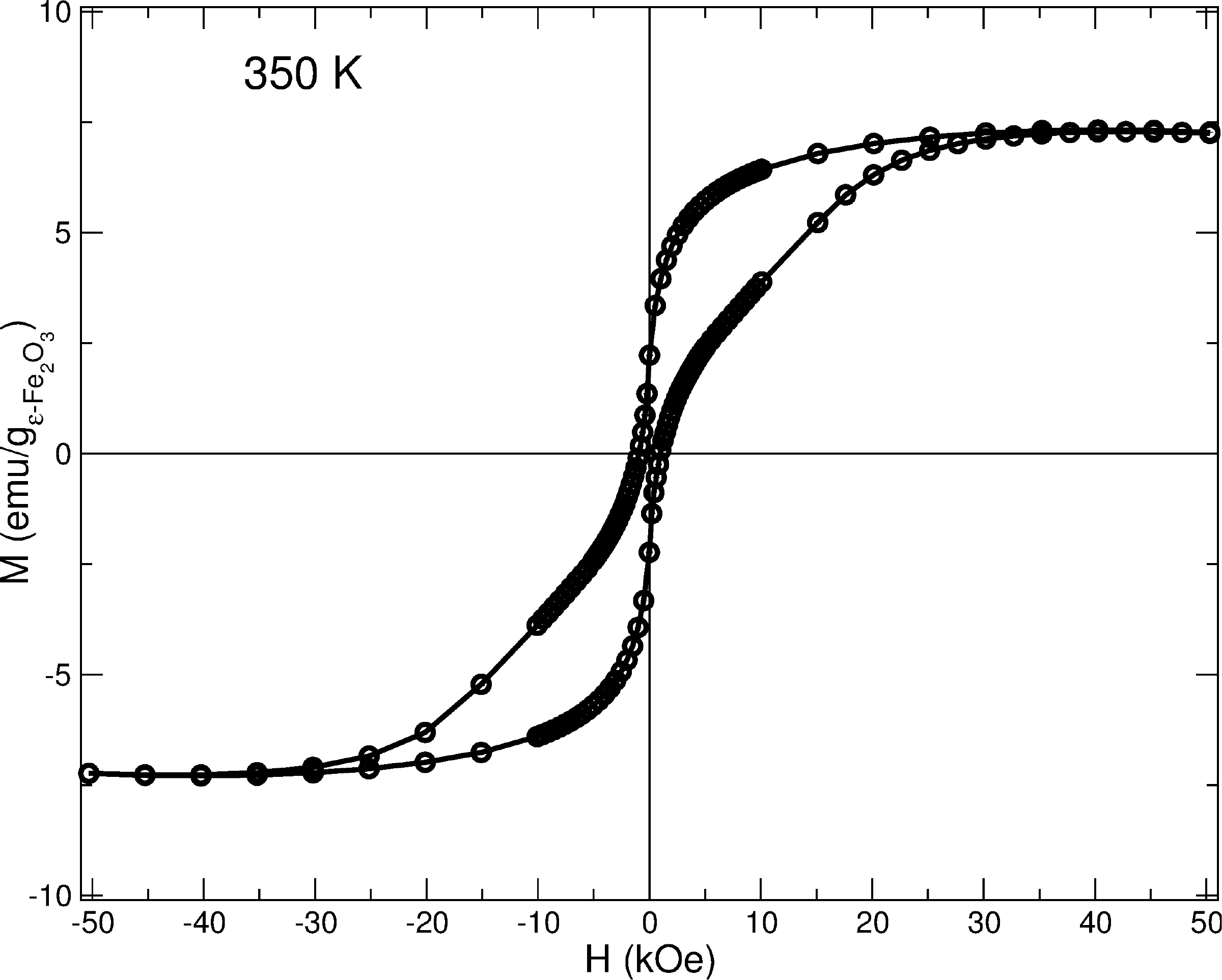}}

\caption{Respective field-dependent magnetizations of the $\epsilon$-Fe$_2$O$_3$ nanoparticles at a) 2~K, b) 100~K, c) 200~K and d) 350~K.\label{fig:MvsH}}

\end{figure*}

Low field AC magnetic susceptibility measurements present magnetism that is in keeping with the high-temperature to low-temperature (meta)magnetic transition of $\epsilon$-Fe$_2$O$_3$\cite{Gich.2006, Sakurai.2005}.   Figure~\ref{fig:Xac} shows the results of zero-field cooled 2.5~Oe AC susceptibility measurements in $\nu$=10~Hz -- 1~kHz.  With warming from 2~K, the in-phase signal, $\chi_{\rm AC}'(\nu,T)$ increases in a frequency independent manner quickly until 25~K, whereupon there is a divergence of the frequency responses, where lower frequencies (slower measuring time) $\chi_{\rm AC}'(T)$ yield larger susceptibilities, but the temperature dependencies of $\chi_{\rm AC}'(\nu)$ are essentially identical with a plateau between $\sim$35 and 75~K.  Above 75~K, there is a further increase in $\chi_{\rm AC}'(T)$ with a peak at 105$\pm$2~K, in agreement with the magnetic transition of the $\epsilon$-phase.  While the peak temperature is frequency independent (Fig.~\ref{fig:Xac}), the amplitude of $\chi_{\rm AC}'(\nu,T)$ decreases quite significantly with increasing measuring frequency, indicating that the spin reorientation is occurring over a small range of time scales, with the magnetic fluctuations (response of the in-phase component) falling out of the measuring time window.  Above 105~K, there is a rapid decrease in $\chi_{\rm AC}(\nu,T)$ with some weak frequency dependence that disappears by 250~K, reflecting the onset of spin dynamics, which the M\"ossbauer spectra collected over that temperature range reveals (discussed below), and likely from previously frozen frustrated spins (e.g. at the Td sites)\cite{Xu.2018} thawing with warming.  The high temperature $\chi_{\rm AC}'(T)$ behaviour is similar to previous results on 25~nm $\epsilon$-Fe$_2$O$_3$ nanoparticles\cite{Gich.2006} with a peak occurring at the transition around 100~K.  However, the second sharp maximum in $\chi_{\rm AC}(T)$ at $\sim$90~K present in 25~nm nanoparticles has been suppressed to $\sim$35~K (with the plateau ending by 75~K) indicating that the shift in magnetic structures has been driven to lower temperatures in these 8~nm nanoparticles.  In keeping with this, the dissipative, out-of-phase component ($\chi_{\rm AC}''(T)$) are qualitatively identical with maxima at 90 and 100~K for the 25~nm crystallites and 25$\pm$2~K and 110$\pm$5~K for these 8~nm crystallites (inset to Fig.~\ref{fig:Xac}).  Both $\chi_{\rm AC}''(\nu,T)$ and $\chi_{\rm AC}'(\nu,T)$ show the same frequency-independent, amplitude-dependent temperature trends.

Figure~\ref{fig:chiDC} shows the low field DC susceptibility ($\chi_{\rm DC}(T)$) temperature dependence in 100~Oe and 1~kOe fields.  Results are qualitatively similar to larger 25~nm spheres\cite{Gich.2006} and 20$\times$100~nm rods\cite{Tadic.2017} of $\epsilon$-Fe$_2$O$_3$.  However, like the $\chi_{\rm AC}(\nu,T)$ discussed above, $\chi_{\rm DC}(T)$ indicates atypical nanomagnetism for iron-oxide single-domain nanoparticles.  Consider the 10~mT temperature scans.  Once the $\epsilon$-Fe$_2$O$_3$ nanoparticles were cooled to 2~K in zero field, with warming $\chi_{\rm ZFC}(T)$ increases until a plateau between $\sim$30 and 75~K (Fig.~\ref{fig:chiDC} bottom), followed with a further significant increase with warming to $\sim$105~K.  These features are linked to the thermal evolution of the magnetic structure going from its low temperature to high temperature configuration\cite{Gich.2006, Sakurai.2005, Tadic.2017}.  With further warming, between 100 and 150~K, there is a slight increase in $\chi_{\rm ZFC}(T)$ that is hinting at the effects of small distortions occurring between the Fe-O bonds\cite{Gich.2006}, altering the magnetism in a manner similar to the Verwey transition in Fe$_3$O$_4$\cite{Attfield.2012}; the temperature evolution of the site-specific bonding and magnetism discussed below will provide more insights.  Between 150 and 400~K, $\chi_{\rm ZFC}(T)$ decreases gradually as thermally driven fluctuations reduce the nanoparticles' magnetizations over the measurement time.  With cooling from 400~K in the 100~Oe field, $\chi_{\rm FC}(T)$ is less than $\chi_{\rm ZFC}(T)$, atypical behaviour for a nanomagnetic system.  From first principles calculations\cite{Ahamed.2018} $\epsilon$-Fe$_2$O$_3$ should have a tetrahedral (Td, Fe$_4$-site) based high anisotropy component, and these sites are frustrated with a strong temperature dependent site magnetization (identified previously\cite{Gich.2006, Ohkoshi.2013, Tseng.2009} and characterized below).  It is likely that the 100~Oe field may not be able to fully couple the the overall ferrimagnetic spin structure resulting in the observed irreversibility; similar behaviour has been reported previously\cite{Zhao.2006, Kollu.2014}.

With cooling, $\chi_{\rm FC}(T)$ presents a `bump' between $\sim$150 and 115~K, the established signature of the onset of the transition in the $\epsilon$ phase\cite{Sakurai.2005}, and $\chi_{\rm FC}(T)$ and $\chi_{\rm ZFC}(T)$ overlap until $\sim$115~K, where the low temperature structure is established.  Below this temperature,  $\chi_{\rm FC}(T)$ and $\chi_{\rm ZFC}(T)$ diverge, tracking with the irreversible magnetization processes identified with $\chi_{\rm AC}''(\nu,T)$ discussed above.  Using a larger measuring field (1~kOe shown in Fig.~\ref{fig:chiDC}) suppresses the signatures of the transitions discussed above, as the larger field quenches the signatures of the changes in magnetic structure with warming and cooling and the spin frustration (e.g. above $\sim$150~K, $\chi_{\rm FC}(T)$ and $\chi_{\rm ZFC}(T)$ overlap  -- see $\Delta \chi_{\rm FC-ZFC}(T)$ in the inset of Fig.~\ref{fig:chiDC}), and the frustrated magnetism from the temperature dependent higher anisotropy Fe$_4$ Td site is no longer observable\cite{Sakurai.2005, Xu.2018}. 

Figure~\ref{fig:MvsH} shows representative 400~K zero-field cooled hysteresis loops between 2 to 350~K that show changes to the overall magnetism of the nanoparticles:  The loop shapes change quite dramatically between temperatures, providing a measure of the impact of the spin reorientation transition and the complexity of the thermal evolution of the overall magnetic anisotropy.  The loops hint at an interesting interaction between the Fe$_1$-to-Fe$_4$ sites' electronic structure and magnetism, examined in detail using M\"ossbauer and x-ray synchrotron spectroscopies, discussed later.

\begin{figure}[t!]

\includegraphics[scale=0.275]{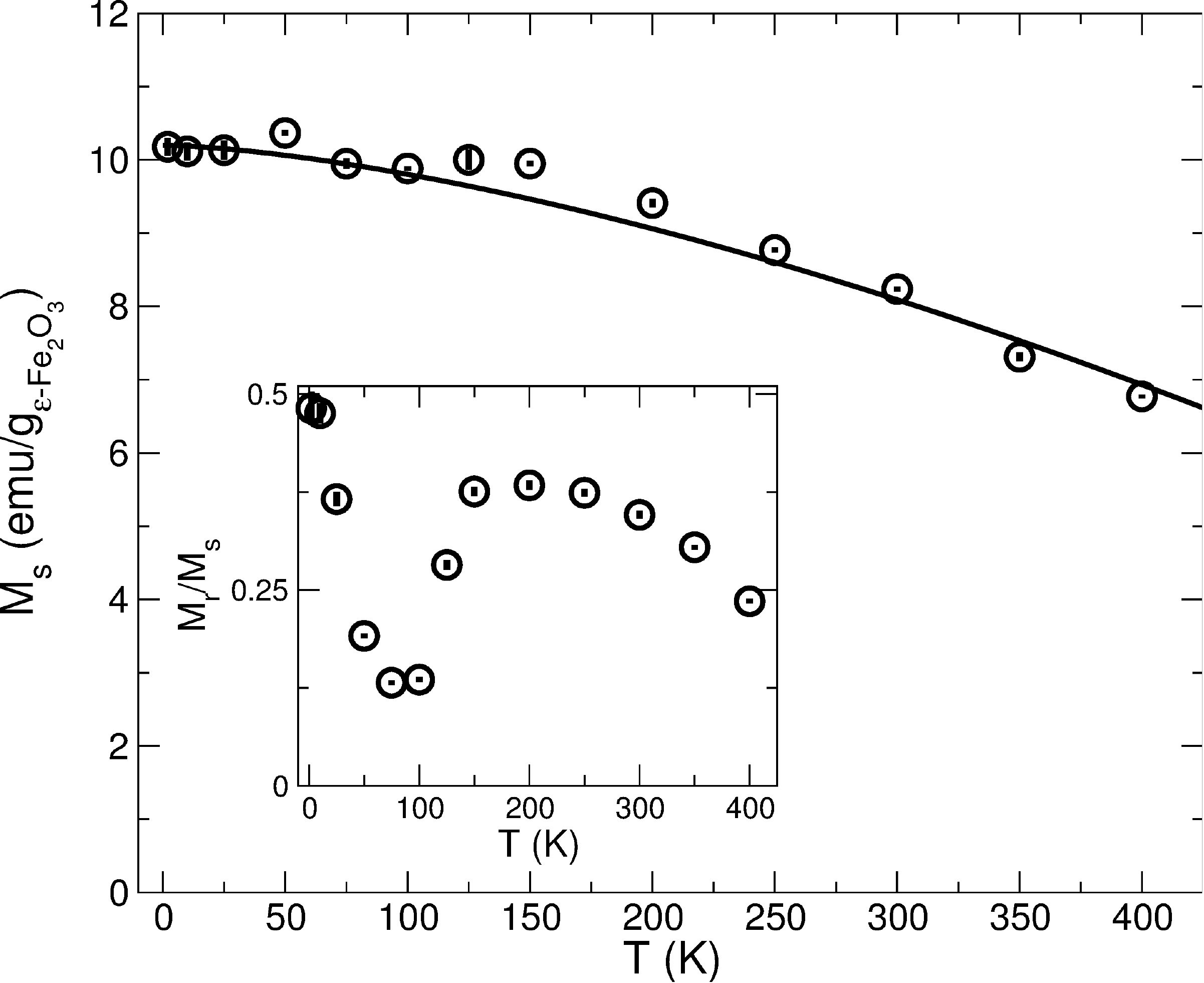}

\caption{Temperature dependence of the saturation magnetization $M_s (T)$ for the $\epsilon$-Fe$_2$O$_3$ nanoparticles.  The solid line is a Bloch $T^{3/2}$-law fit as described in text.  The inset shows the temperature dependence of the remanent magnetization ($M_r$) normalized to $M_s$ at that temperature.\label{fig:MsvsT}}

\end{figure}

Figure~\ref{fig:MsvsT} presents the temperature dependence of the saturation magnetization ($M_s$) and remanent magnetization ($M_r$) of the $\epsilon$-Fe$_2$O$_3$ nanoparticles.  $M_s(T)$ is well described over the complete range of temperatures using the Bloch T$^{3/2}$ law for collective spin oscillations, $M_s(T)/M_s(2~K) =(1-BT^{3/2})$, where $B=3.53\pm 0.04\times 10^{-5}$ (solid line in Fig.~\ref{fig:MsvsT}) in good agreement with the Bloch constant of similar sized iron oxide nanoparticles\cite{Goya.2003, Barbeta.2010}.  Interestingly, $M_s$ separates from the $T^{3/2}$ trend at 50~K and between 125 and 200~K, which indicates that the spin wave behaviour is affected by the transition from low temperature to high temperature magnetic phases and the concomitant anisotropy changes.  The temperature dependence of $M_r/M_s$ (inset to Fig.~\ref{fig:MsvsT}) shows the impact of these changes even more clearly, especially below 200~K, in agreement with previous results on $\epsilon$-Fe$_2$O$_3$ nanoparticles\cite{Tseng.2009, Gich.2006}.

\begin{figure}[b!]

\includegraphics[scale=0.33]{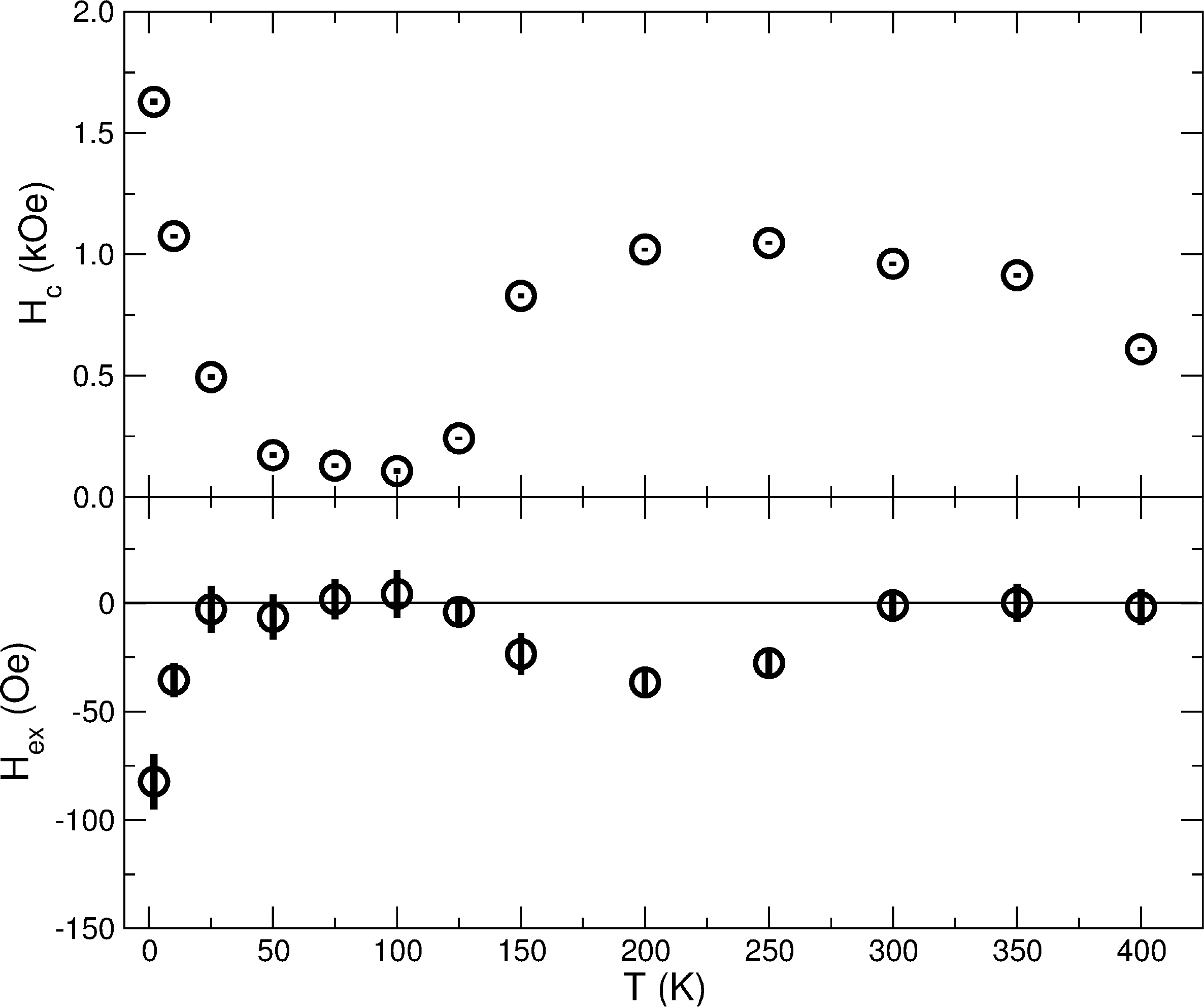}

\caption{Temperature dependence of the $\epsilon$-Fe$_2$O$_3$ nanoparticle coercivity ($H_c$, top) and the exchange bias field ($H_{ex}$, bottom).\label{fig:HcHex}}

\end{figure}

The unusual temperature dependence of the $\epsilon$-Fe$_2$O$_3$ nanoparticles' remanent magnetization ($M_r$) from the hysteresis loop measurements (e.g.~Fig.~\ref{fig:MvsH}) is shown clearly by $M_r/M_s(T)$ in the inset of Fig.~\ref{fig:MsvsT}.   At the lowest temperatures (2 -- 10~K) $M_r/M_s$$\sim$0.5, in keeping with a random distribution of noninteracting (i.e. no dipolar effects) uniaxial single domain nanoparticles\cite{Stoner.1948}.  With warming into the transition from 25 to 100~K there is a significant decrease of $M_r/M_s$ to $\sim$0.15, which is indicating that the nanoparticles' spin (domain) structure is being altered, e.g. the ferrimagnetic spin configuration is changing with warming as is the magnetocrystalline anisotropy.  With further warming to $\sim$150~K, where the high temperature phase takes a hold of the magnetic structure, there is an equally rapid return of $M_r/M_s$$\sim$0.4 that decreases monotonically to $\sim$0.25 by 400~K due to thermal effects and the temperature dependence of the anisotropy ($M_r/M_s \propto f(\Delta E) \Delta T$ where $\Delta E = K V/k_B T$). 

The shifting of spin configurations of the $\epsilon$-Fe$_2$O$_3$ nanoparticles and the impact on the magnetic anisotropy ($K$) with temperature is most unmistakably observed in the coercivity's temperature dependence ($H_c(T)=(H_{c1}-H_{c2})/2$ where $H_{c1}$ and H$_{c2}$ are the negative field and positive field coercivities, respectively) presented in Fig.~\ref{fig:HcHex}.  From 400 to $\sim$150~K, $H_c(T)$ is weakly temperature dependent, but with the onset of the transition with further cooling, $H_c(T)$ rapidly decreases to $\sim$0 until the onset of the low temperature magnetic structure is established at $\sim$50~K.  Once the low temperature spin configuration has been frozen in (as revealed explicitly by the temperature dependence of the M\"ossbauer spectroscopy and XMCD site-specific magnetism discussed below) $H_c(T)$ increases monotonically to its maximum value at 2~K.

At least 50\% of the Fe$^{3+}$ ions are on or near the surface (a unit-cell thick) of the 8~nm crystallites.  Also, the Fe$_1$ and Fe$_2$ Oh and Fe$_4$ Td sites experience disorder even inside the `core'.  Thus, a considerable number of the Fe$^{3+}$ ions should be magnetically frustrated (e.g. will experience broken bonds or are uncompensatated, which we quantify using M\"ossbauer spectroscopy).  This should result in the observed complex hysteresis loop shapes due to the competing surface and core anisotropies, and also can enable a unidirectional anisotropy that results in an exchange bias loop shift -- behaviour that has been observed in similar sized $\gamma$-Fe$_2$O$_3$ nanoparticles\cite{Skoropata.2014}, and many other ferrite-based nanomagnets, including core/shell structured magnets\cite{Skoropata.2017}.  We observed a small but measurable `spontaneous' exchange bias loop shift ($H_{ex}(T)=(H_{c1}+H_{c2})/2$, bottom panel of Fig.~\ref{fig:HcHex}).  $H_{ex}(T)$ mirrors the results of the spin reorientation transition between 25~K and 150~K (as identified with the $\chi_{\rm AC}(\nu,T)$ and the site-specific magnetism discussed below) and we link this exchange coupling to both spin frustration and the spontaneous symmetry breaking of the magnetic transition ($H_{ex}(T)$$\sim$0 through the high-temperature to low-temperature spin transition).  Similar magnetism has been reported in the structural/magnetism multiferroic analogue BFO\cite{Dong.2011, Maity.2013}.  Field-cooling from 400~K in a 50~kOe field removes $H_{ex}$ and leaves $H_c$ unaffected.  We postulate that the 50~kOe field during cooling (especially through the spin transition) forces the alignment of the spins so as to weaken or entirely remove the exchange coupling pathways that enable $H_{ex}$.

\begin{figure*} 

\subfigure[\ ]{\includegraphics[scale=0.25]{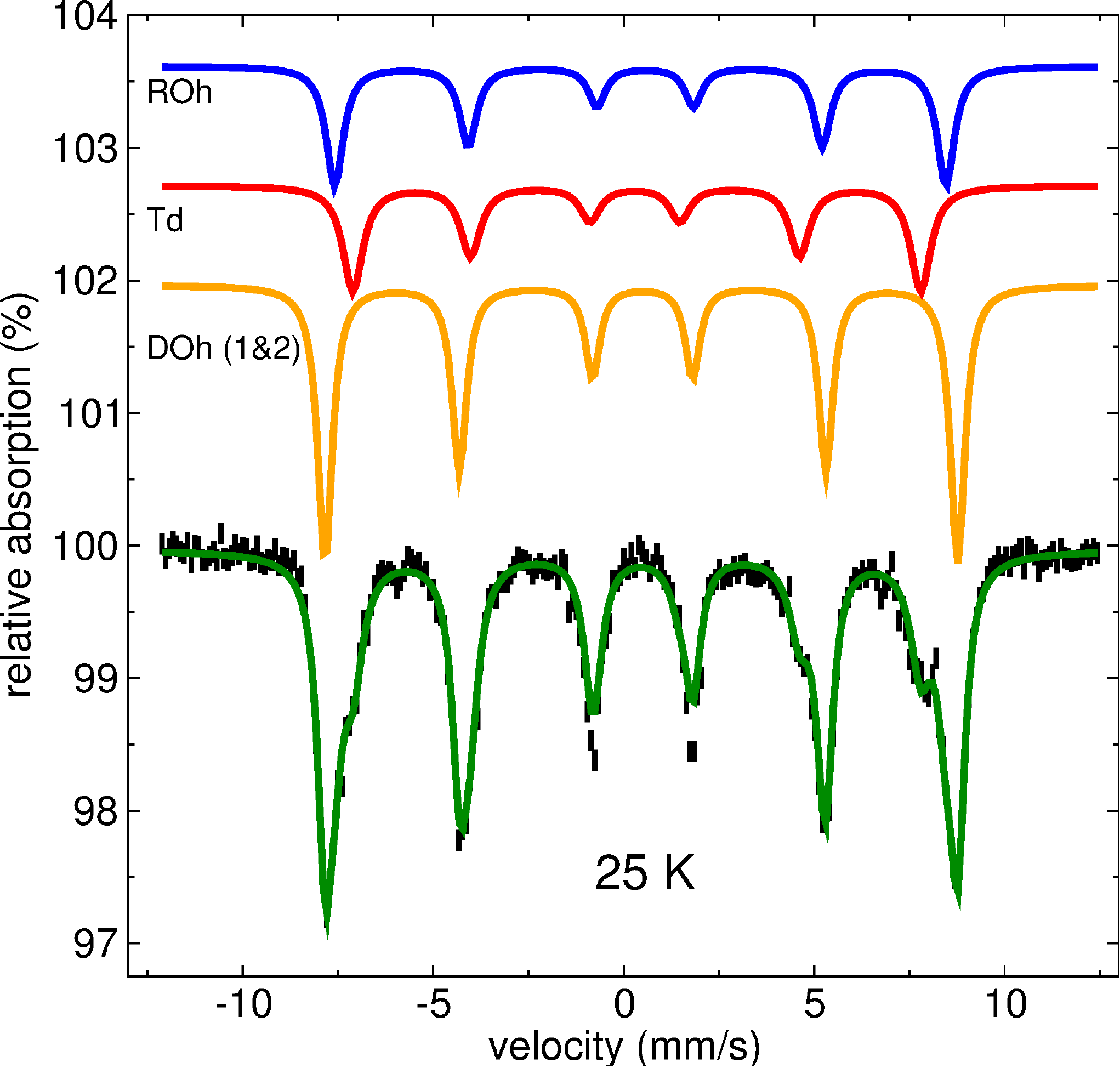}}
\subfigure[\ ]{\includegraphics[scale=0.25]{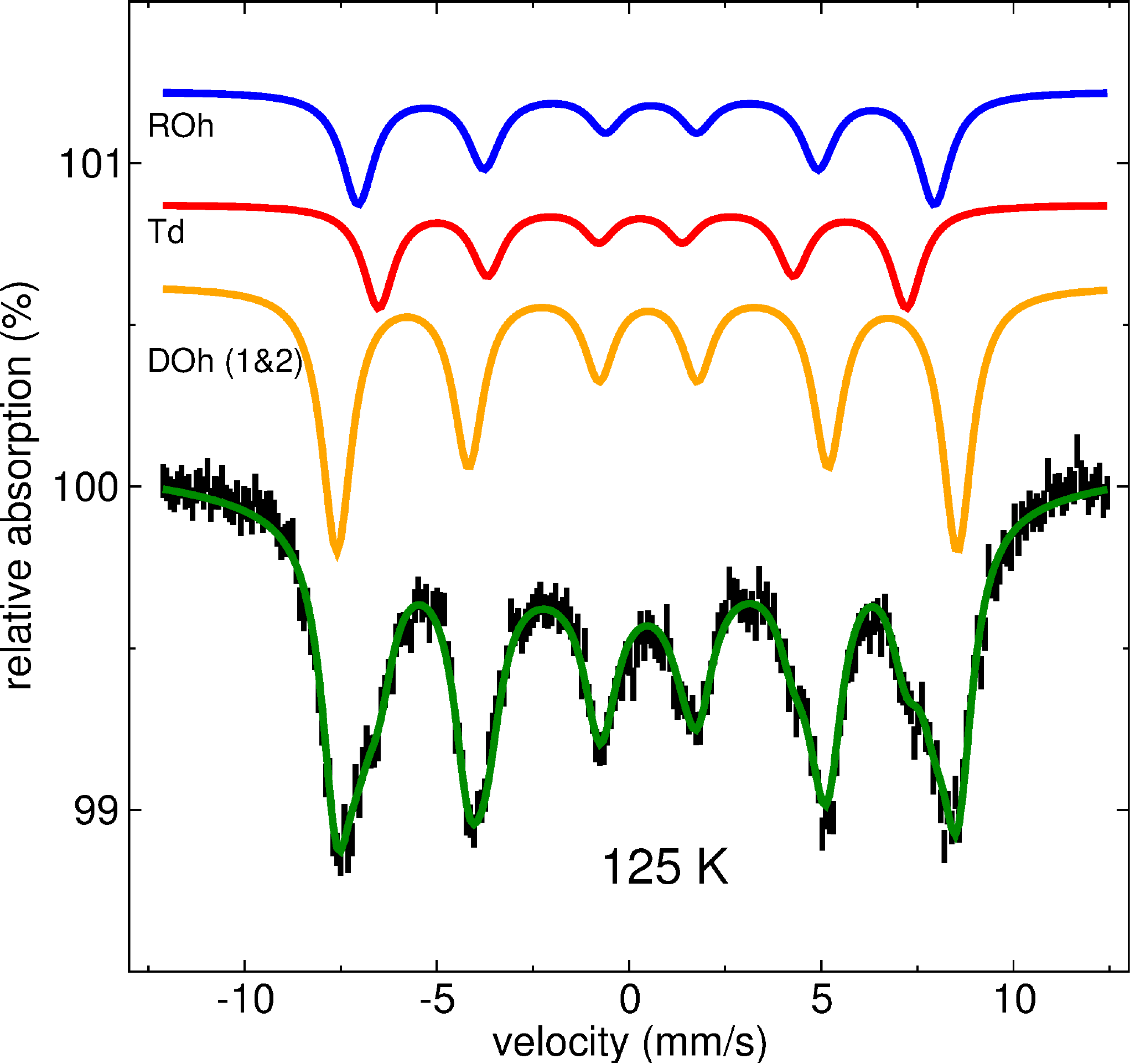}}
\subfigure[\ ]{\includegraphics[scale=0.25]{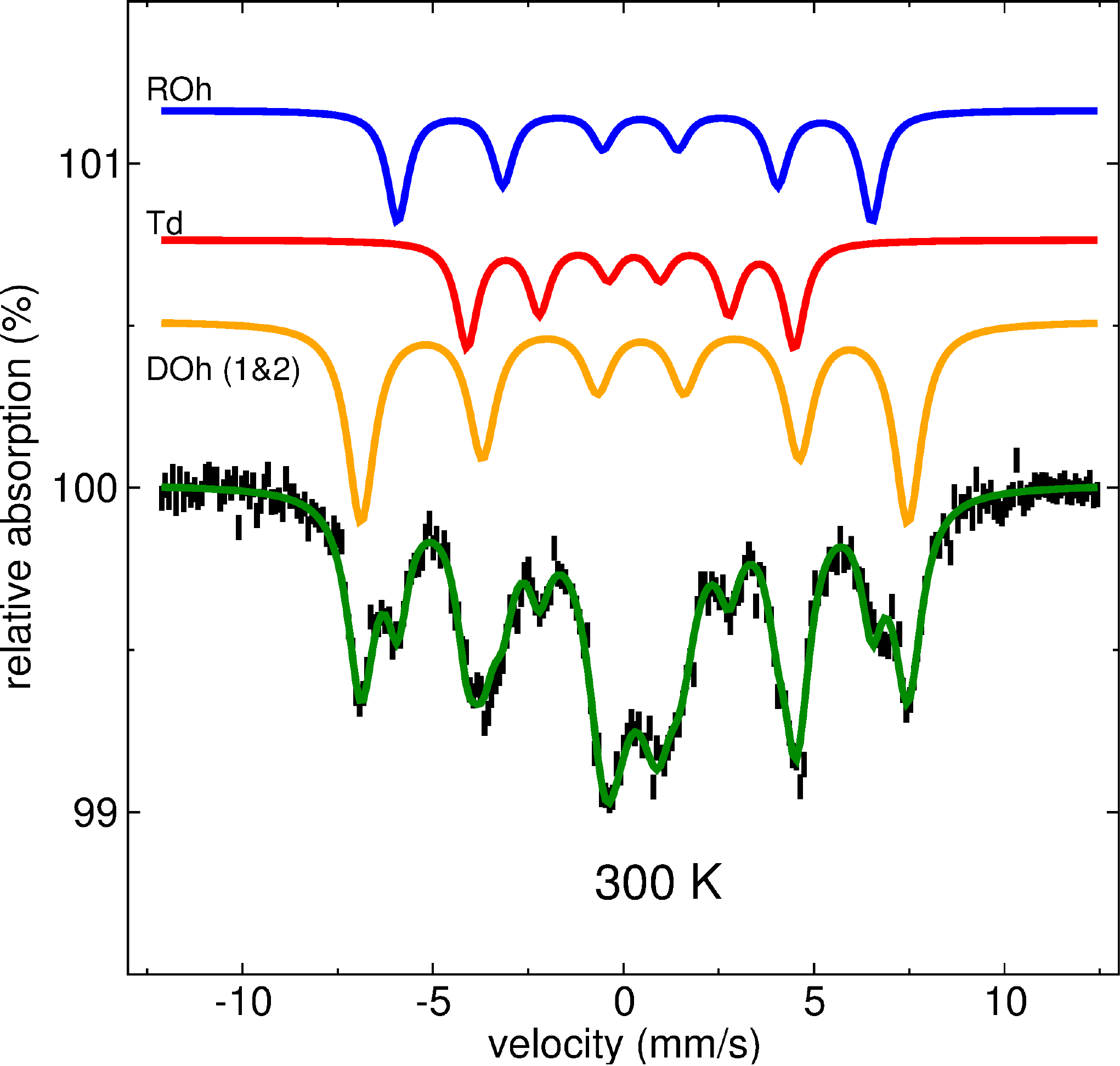}}

\caption{(colour online) Selected M{\"o}ssbauer spectra at a) 25~K, b) 125~K and c) 300~K, with the green solid line the full component fit and the blue, red and orange solid lines present the site-assigned spectral decompositions, as discussed in the text.\label{fig:Mossy}}

\end{figure*}

\begin{figure}[b!]

\includegraphics[scale=0.33]{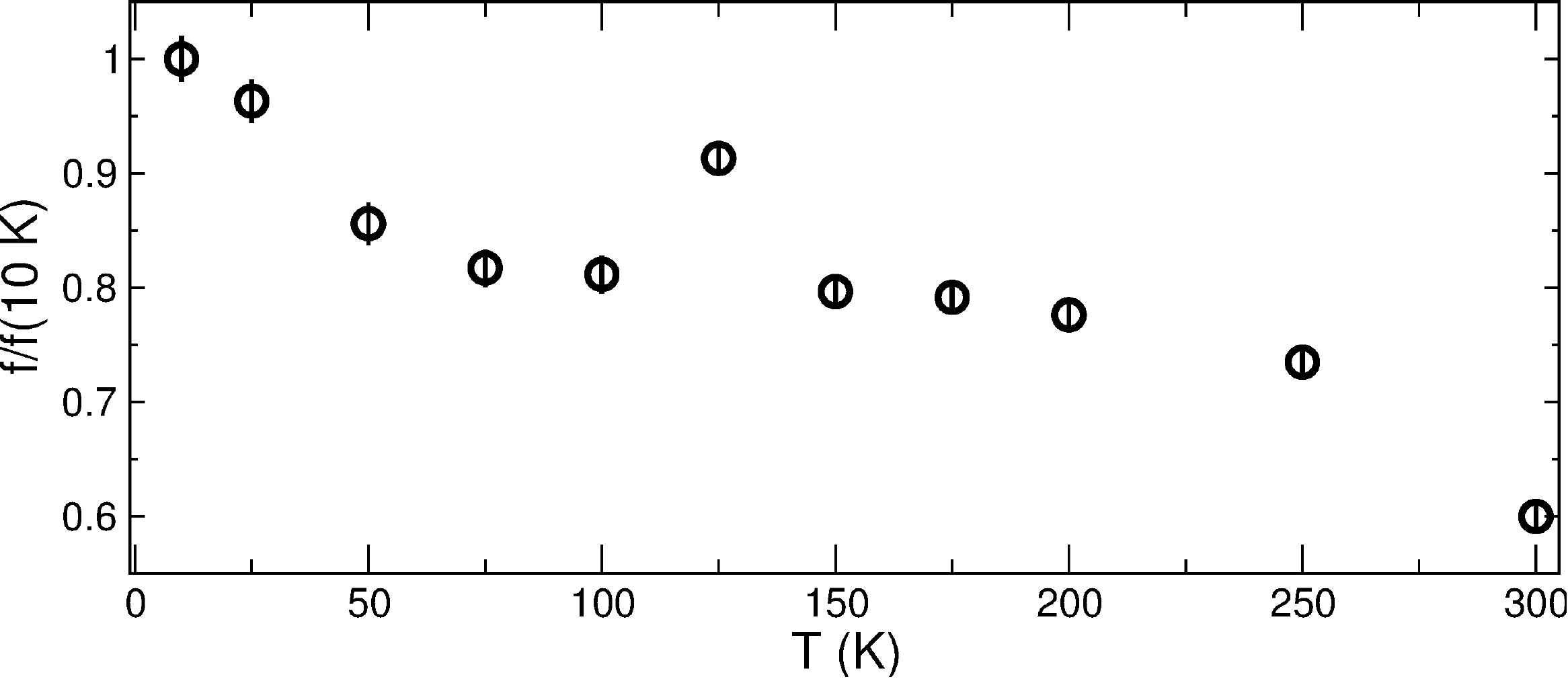}

\caption{Temperature dependence of the M\"ossbauer spectra total relative absorption, $f$ (Debye-Waller factor) for the $\epsilon$-Fe$_2$O$_3$ nanoparticles.\cite{ffactor} \label{fig:ffactor}}

\end{figure}

To clearly characterize the site-specific structure and magnetism changes identified by the magnetometry and susceptometry, we turn to the information provided by M\"ossbauer spectroscopy that measures the atomic $^{57}$Fe magnetism.  M\"ossbauer spectroscopy uses the resonant absorption and recoilless emission of gamma rays to excite transitions in the probe nuclei.  The atomic-level electronic and magnetic environments of $^{57}$Fe atoms throughout the sample volume are characterized by the hyperfine parameters that affect the energy required to excite the transitions.  These transitions are identified by the hyperfine field ($B_{\rm hf}$) that describes the magnetic environment, and the isomer shift ($\delta$) and quadrupolar splitting ($\Delta$) that reflect the local electronic environment about the $^{57}$Fe nucleus in the nanoparticles, and the spectral component linewidths ($\Gamma$) due to the lifetime of the excited state of the nucleus, with any measured increase in $\Gamma$ compared to that of the 6~$\mu$m $\alpha$-Fe foil related to structural or chemical disorder from a (small) distribution of hyperfine parameters away from the site-specific values (e.g. of the ideal, bulk structure).  Example spectra at temperatures through the low temperature to high temperature regimes of the spin transition are shown in Fig.~\ref{fig:Mossy}.  The temperature dependence of the spectral total relative absorption, i.e. the $^{57}$Fe recoil free fraction ($f$-factor, or Debye-Waller factor), presented in Fig.~\ref{fig:ffactor}, clearly identifies the change in the magnetic configuration starting at $\sim$150~K and evolving with further cooling.  The $f$-factor is related to the crystal lattice phonon-modes and is very sensitive to changes in electronic bonding (e.g. the onset of a magnetic order at $T_C$)\cite{Gutlich.2010}, and the increase in $f(T)$ at $\sim$125~K  is in agreement with the magnetic transition detected with neutron diffraction\cite{Gich.2005, Gich.2006}.

\begin{figure}[b!]

\subfigure[\ ]{\includegraphics[scale=0.165]{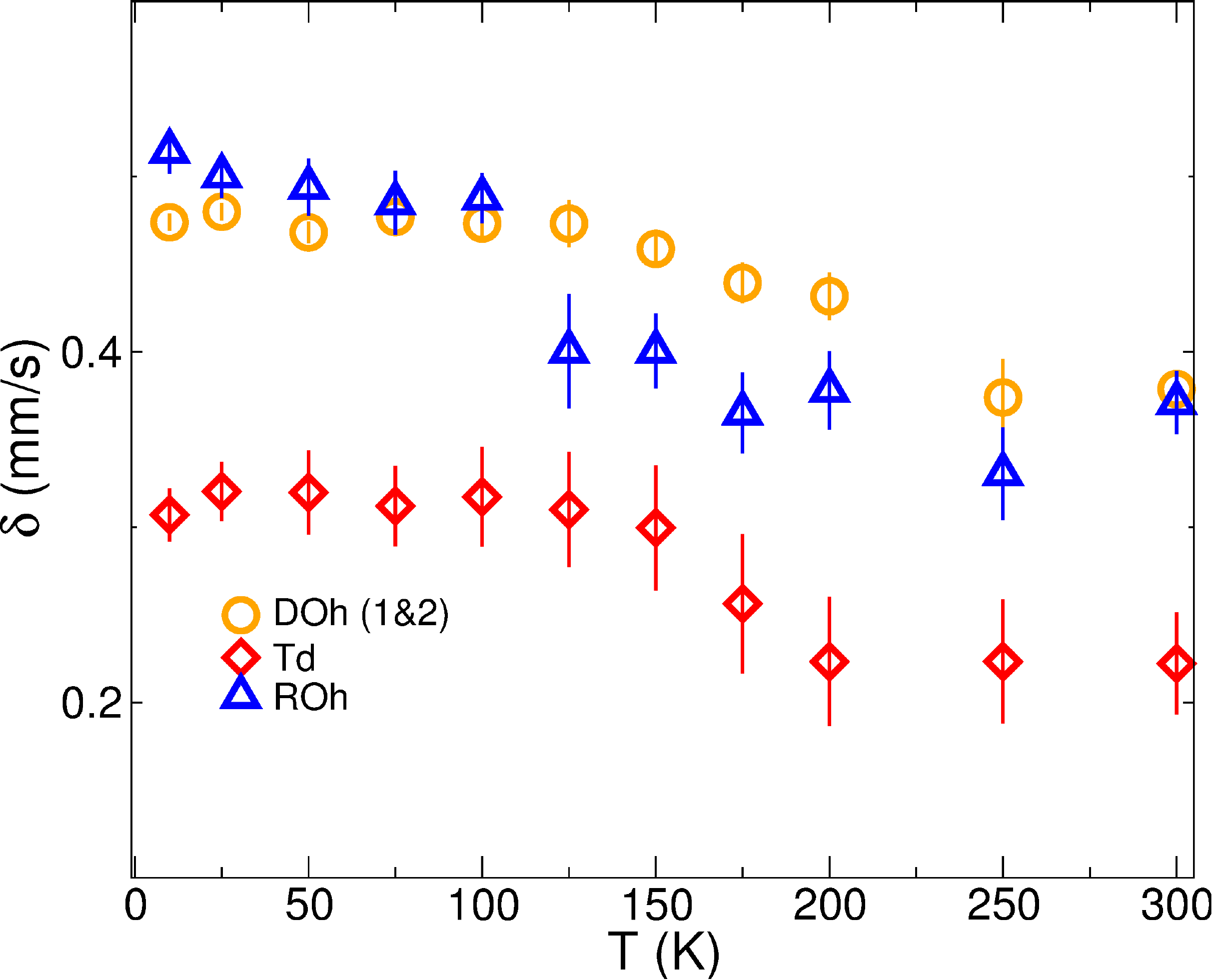}}
\subfigure[\ ]{\includegraphics[scale=0.165]{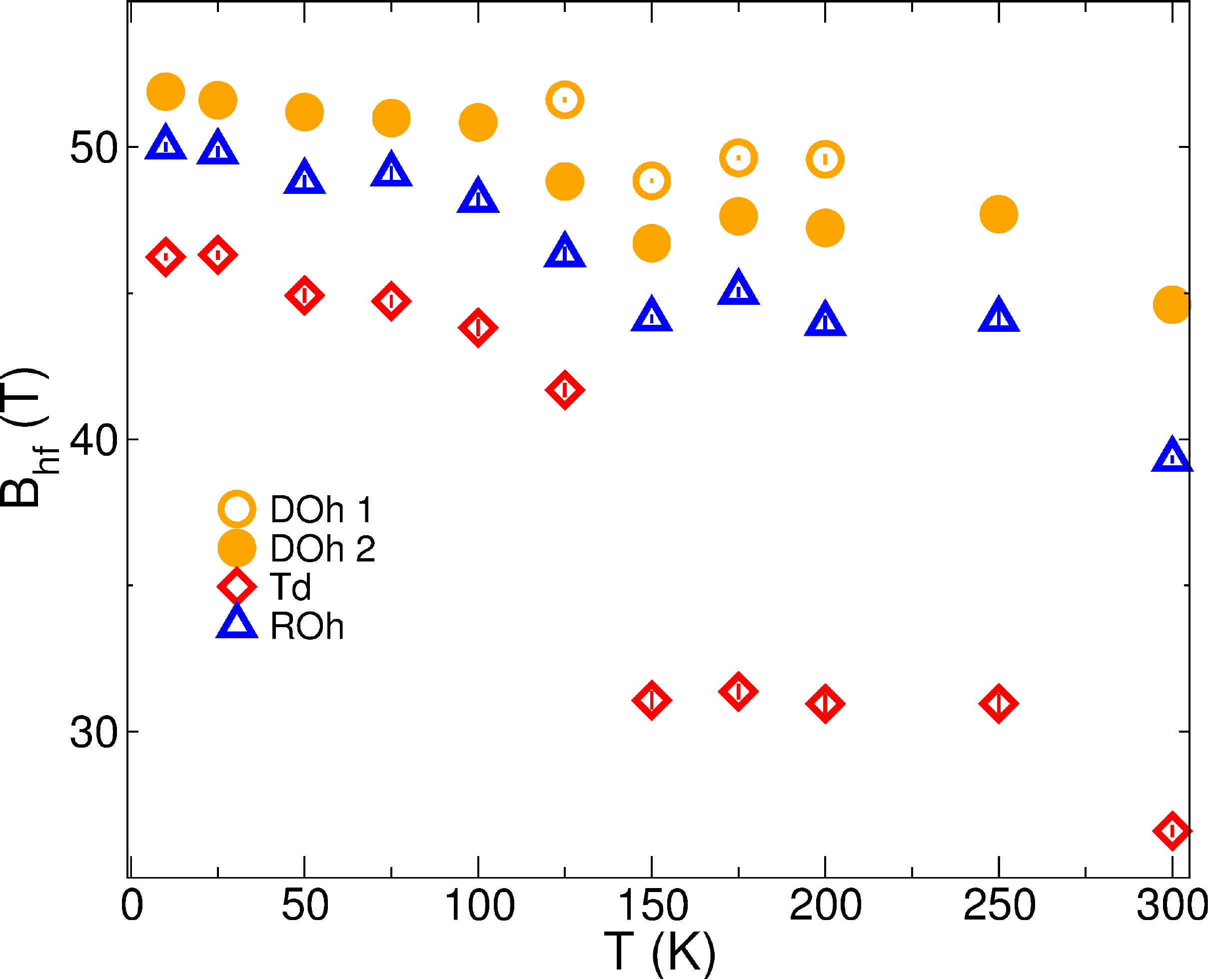}}

\subfigure[\ ]{\includegraphics[scale=0.165]{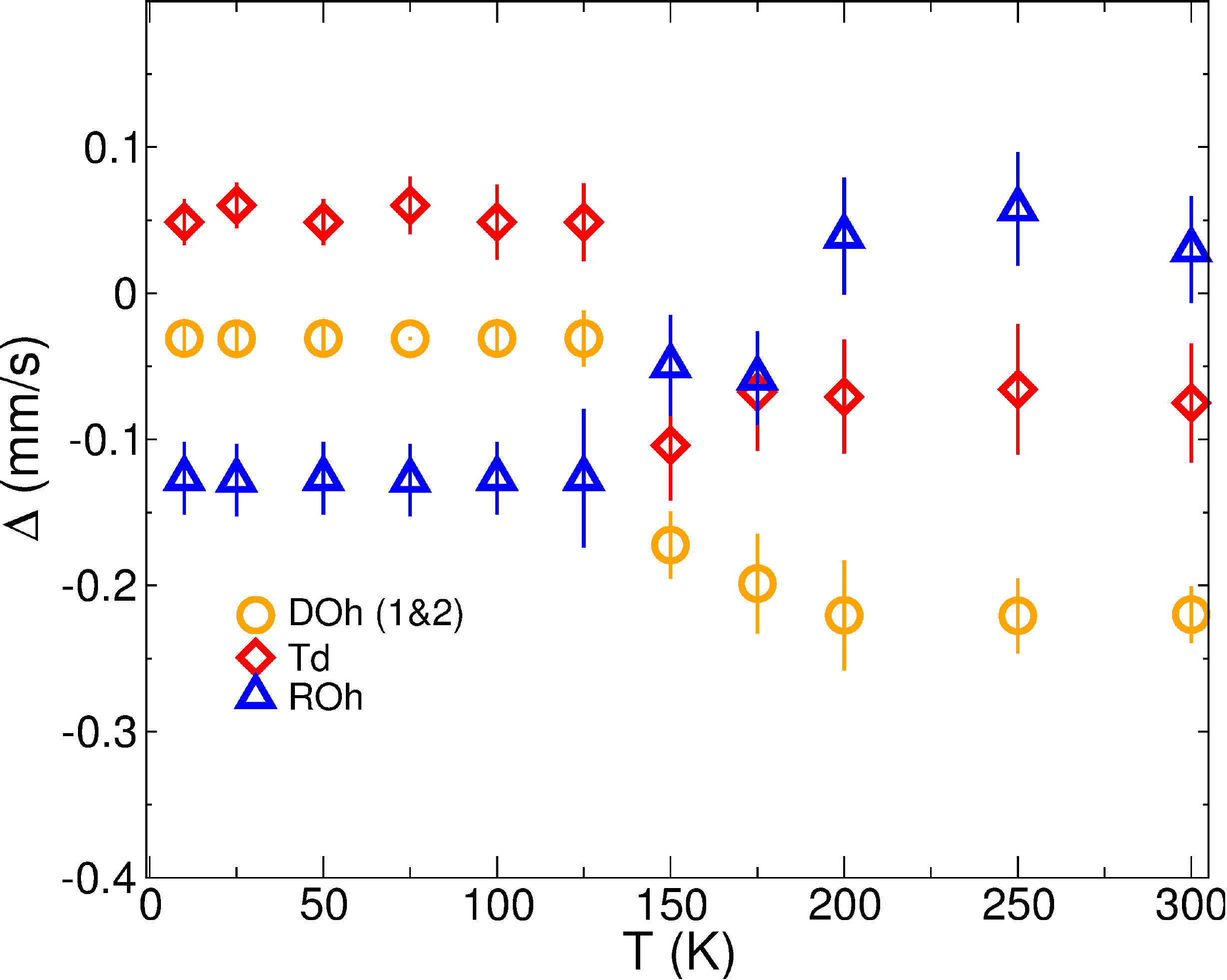}}
\subfigure[\ ]{\includegraphics[scale=0.165]{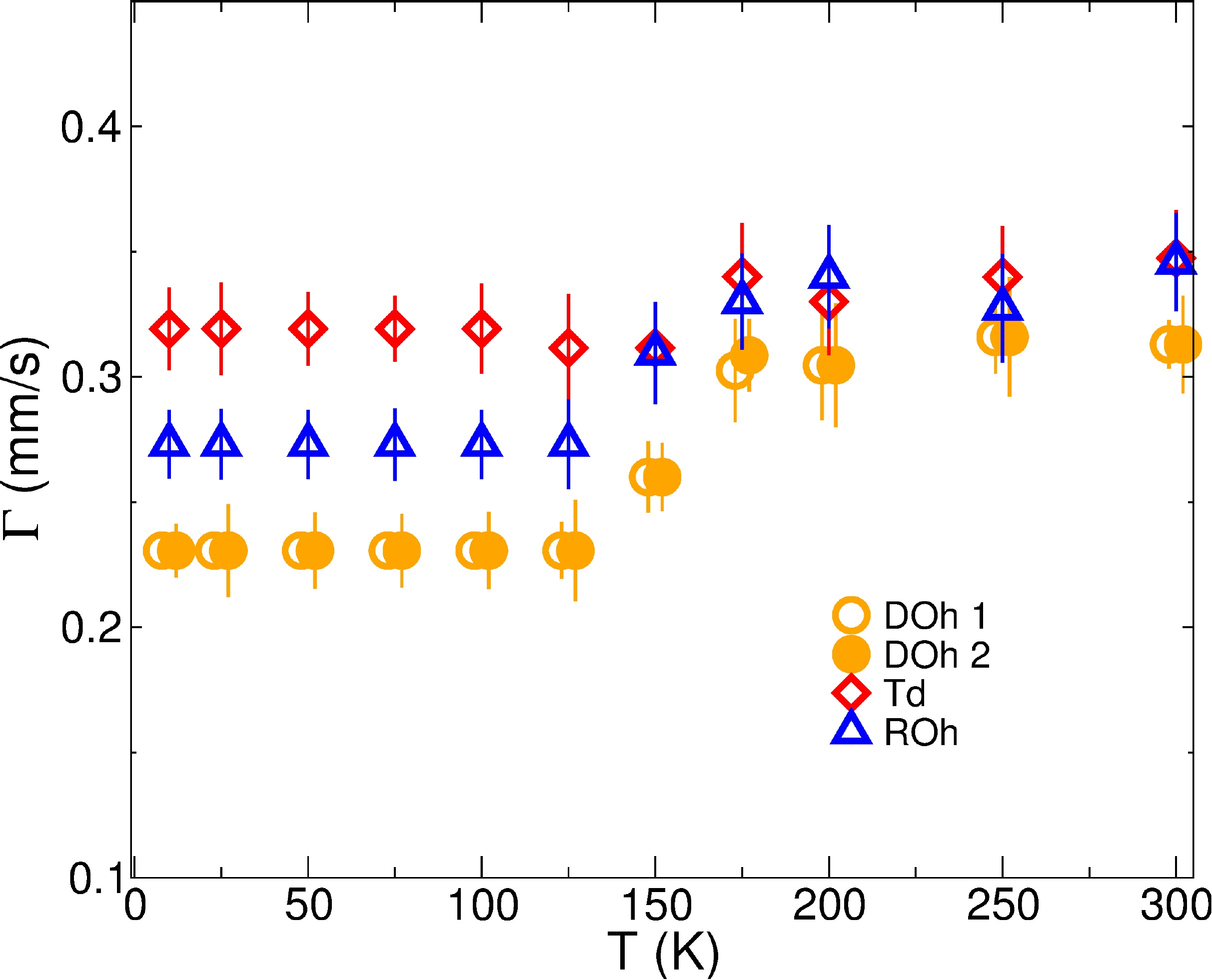}}

\caption{(colour online) Temperature evolution of the hyperfine parameters determined from fits to the M\"ossbauer spectra.  a) Isomer shifts, $\delta$, b) hyperfine fields, $B_{\rm hf}$, c) quadrupolar shifts, $\Delta$, and linewidths, $\Gamma$, of the disorted octahedral (DOh), regular octahedral (ROh) and tetradedral (Td) Fe sites.  \label{fig:mossyfits}}

\end{figure}

Figure~\ref{fig:mossyfits} shows the fit results to the M{\"o}ssbauer spectra.  Overall, the interesting site-specific behaviour presented agrees with previous observations\cite{Gich.2006}, including the general trends observed in the hyperfine fields and isomer shifts.  Since these $\epsilon$-Fe$_2$O$_3$ nanoparticles are significantly smaller, the spectra exhibit the effects of structural and chemical disorder due to finite-size effects (e.g. spectral linewidths significantly broader than the nature linewidth of the $\gamma$-ray source (0.133$\pm$0.004~mm/s)).  Four components for the different Fe sites were used\cite{relaxfit}, however both distorted octahedral sites (DOh; Fe$_1$ and Fe$_2$) settled towards the same hyperfine parameters with equivalent $\Gamma$s, with both sites becoming clearly resolvable (e.g. specific $B_{\rm hf}$s for each site) across the transition.  Fig.~\ref{fig:mossyfits}a shows that the isomer shifts of the two distorted octahedral sites have the expected thermal evolution due to second-order Doppler shift effects\cite{Gutlich.2010}, but the tetrahedral and regular octahedral sites show a step-like behaviour through the transition. This can be ascribed to a decrease in 2$p$ - 3$d$ intermixing, lowering the 4$s$ occupation and allowing a radial decompression of the other $s$-shell charge densities, thereby lowering $|\psi(0)|^2$, the electronic density at the nucleus. A subtle change in the ferrimagnetic structure could precipitate this, via either the direct overlap and covalency of the bonds, or indirectly by a change in the distortion of the coordination polyhedra of the Fe-sublattices, such that the energy separation between the orbitals (through O exchange) is redistributed.  At the low temperatures (10--100~K), the large hyperfine fields (Fig.~\ref{fig:mossyfits}b) observed for all components is typical for high-spin complexes with large Fermi contact fields, and are consistent with $B_{\rm hf}$s of all other iron-oxides; Oh and Td $B_{\rm hf}$s are in good agreement with those of $\alpha$- and $\gamma$-Fe$_2$O$_3$, and Fe$_3$O$_4$\cite{GreenwoodGibb}.  $B_{\rm hf}$ of the Fe$_{1,2}$ DOh sites show that the magnetic transition is `sluggish' with the Fe$_1$ DOh site experiencing changes to the local magnetic environment at a higher temperature than the Fe$_2$ DOh site.  The Fe$_4$ Td site also shows the slow thermal development of the transition, with a slow decrease in $B_{\rm hf 4}$ between $\sim$25 and 100~K (as indicated by e.g. $\chi_{\rm AC}(\nu,T)$ discussed above).  Then, a most dramatic change in local environment through the transition is observed, dropping $B_{\rm hf 4}$ $\sim$40\% in a step-like fashion, that indicated a dominant contribution to the Fermi contact component of $B_{\rm hf}$ via a reduction in core-spin polarization due to a change in charge mixing, and the orbital- and spin-dipolar contributions to the hyperfine field is present.  The change in the local electronic charge distribution symmetry that affects $B_{\rm hf}$ is observable in the step-line change in $\Delta$ values (Fig.~\ref{fig:mossyfits}c) that is common to all sites and consistent with changes in distortion of the coordination polyhedra. The overall changes in hyperfine parameters may also be consistent with a localized-to-itinerant transition in the electronic orbitals, such that the superexchange conditions change spontaneously, e.g.  at the Fe$_4$ sites, and propagate throughout the crystal structure.  The temperature dependence of the fitted linewidths for each site (Fig.~\ref{fig:mossyfits}d) show that the site disorder is constant throughout the transition, specifically indicating a lack of changes in the Fe$_4$ inter-site disorder, and the higher temperature spectra reveal that the amount of disorder about the Fe$_{1,2}$ DOh sites increases slightly with warming through the transition.  In combination, the fit results of the M\"ossbauer spectra show that there is a collective transition centred on the tetrahedral sites, whose effects presumably propagate through to the other Fe sites. 

\begin{figure}[t!]

\includegraphics[scale=0.33]{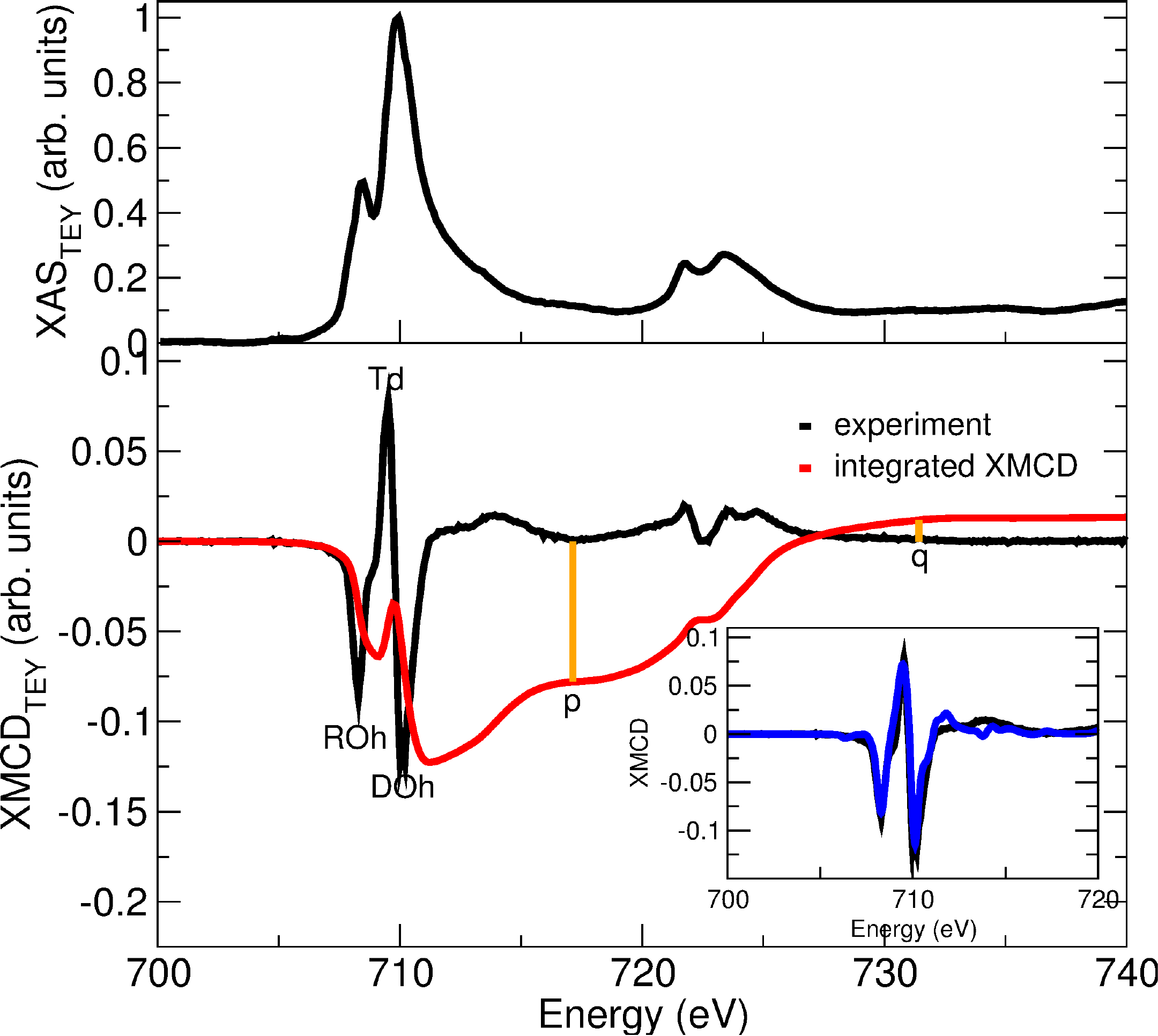}

\caption{(colour online) Representative XAS (top) and artifact-free XMCD (bottom) spectra at 10~K in 5~T.  The red line is the integrated XMCD spectrum, and the energies chosen for $p$ and $q$ are marked, as described in the text.  Inset shows the simulated XMCD spectrum (blue line) of the combined $\sim$(3Fe$^{3+}_{\rm Oh}$ + Fe$_{\rm Td}$) components superposed over the data, described in the text.}
\label{fig:xasxmcd}
\end{figure}

X-ray absorption spectroscopy (XAS) collected over the Fe (700--730~eV) L$_{3,2}$ edges (2$p\rightarrow3d$ transitions) were collected to develop a more detailed picture of the evolution of the Fe electronic configurations with temperature and field (x-ray magnetic circular dichroism, XMCD).  Figure~\ref{fig:xasxmcd} shows the 10~K XAS and XAS normalized XMCD spectra.  The Fe XAS spectrum was typical for a spinel Fe-oxide\cite{BriceProfeta.2005} and is in in good agreement with results on larger $\epsilon$-Fe$_2$O$_3$ nanoparticles\cite{Tseng.2009}.  Ligand-field multiplet simulations\cite{XCLAIM, CTM4XAS} including weighted contributions from all sites (determined from the M\"ossbauer fits) are in good agreement with the spectra (Fig.~\ref{fig:xasxmcd} inset).  The XMCD spectrum site assignment show that the octahedral sites' magnetizations align parallel to each other, and the antiferromagnetically coupled tetrahedral sites align anti-parallel. 

\begin{figure}[b!]
\includegraphics[scale=0.33]{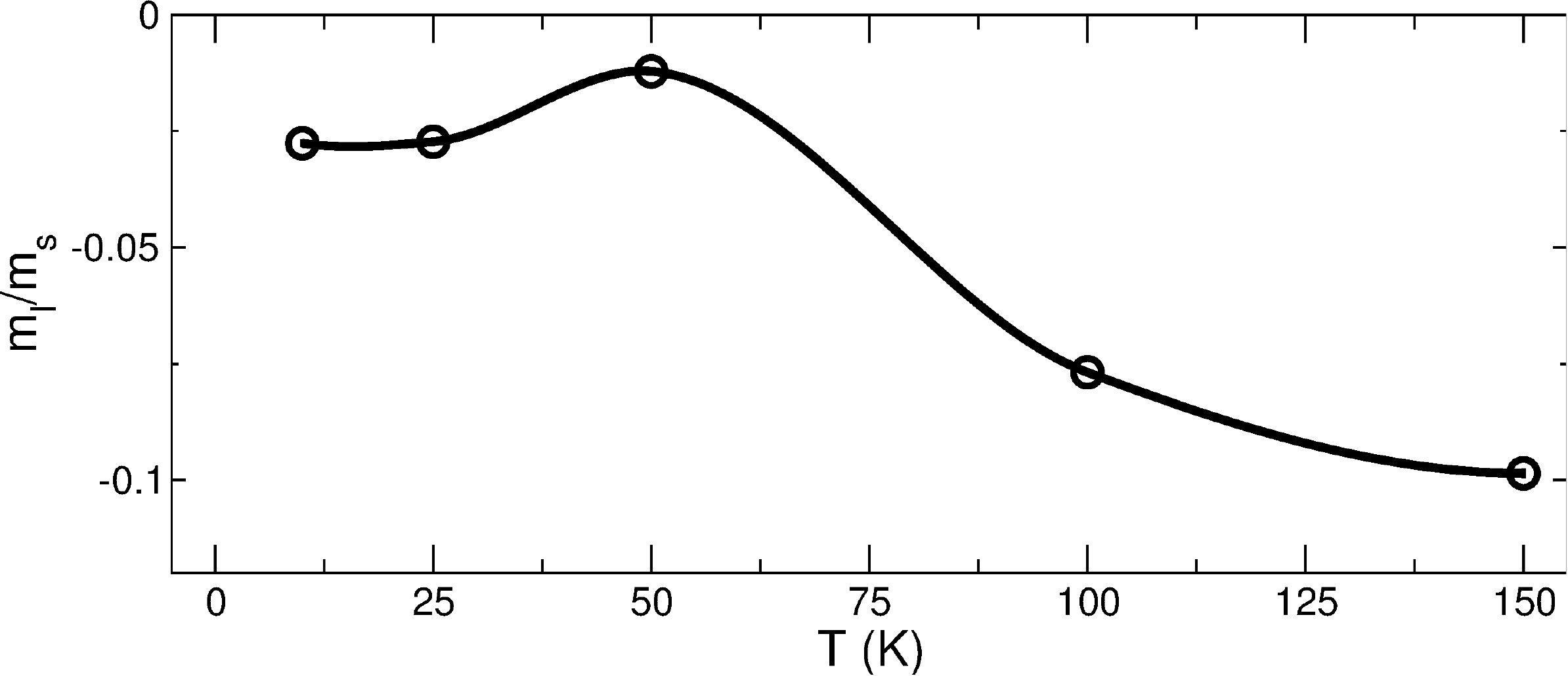}
\caption{Temperature dependence of the orbit-to-spin magnetic moment ratios for the $\epsilon$-Fe$_2$O$_3$ nanoparticles.  The line is a guide to the eye.\label{fig:mo-ms}} 

\end{figure}

Using sum rule analysis\cite{Thole.1992,Carra.1993,Chen.1995} on the XMCD spectra collected at various temperatures, we have determined the relative contributions of spin and orbital moments using $m_{\ell}$/$m_s$=$2q/(9p-6q)$, where $p$ and $q$ are the integrated XMCD intensities over the $L_3$ edge, and the combined $L_3$ and $L_2$ edges, respectively, shown in Fig.~\ref{fig:xasxmcd}.  We find an anti-parallel spin and orbital alignment (the negative sign of $m_{\ell}$/$m_s$), and an increase in $m_{\ell}$/$m_s$ as it enters the spin reorientation transition with warming from 10~K, followed by a rapid decrease of $m_{\ell}$/$m_s$ until 150~K.

\begin{figure}[t!]
\includegraphics[scale=0.33]{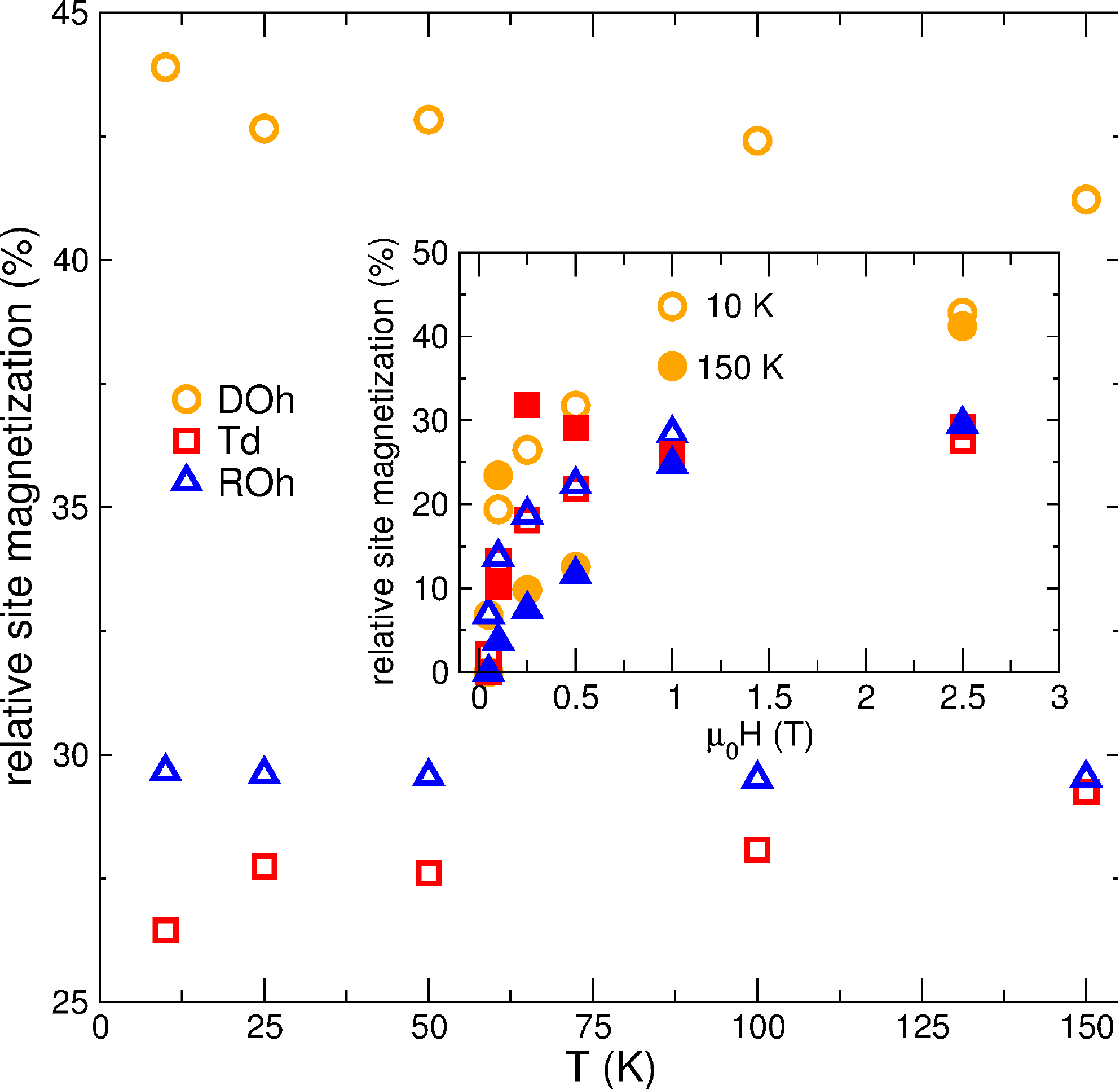}
\caption{(colour online) Relative AF TEY XMCD absorption amplitude versus temperature of the Fe sites from $L_3$ edge analysis.  Inset shows the field dependence of the sites from the $L_3$ edge spectra at 10~K (open symbols) and 150~K (closed symbols).\label{fig:xmcdMvsT}}

\end{figure}

The magnitudes of the relative site magnetization ascertained from the $L_3$ edge amplitudes as a function of temperature is presented in Fig.~\ref{fig:xmcdMvsT}, with the field dependence of the sites at 10 and 100~K presented in the inset.  The temperature dependence of the site magnetizations reflect $B_{\rm hf}(T)$ over the same range of temperatures (keeping in mind that there is no direct relationship between elemental magnetization $M$ and $B_{\rm hf}(T)$ due to complexities of the transferred hyperfine field effects\cite{liz}).  The field dependence of the Fe sites (inset of Fig.~\ref{fig:xmcdMvsT}) identifies a nearly constant magnetization for the tetrahedral site when the field is larger than about 0.5~T. This behaviour, in contrast to the other sites showing a typical $M$ vs $H$ dependence over these field values indicates the alignment of the tetrahedral site occurs at smaller fields than that of the octahedral sites.  This further corroborates the imbalance of the Td Fe$_4$ moments as the source of the ferrimagnetic configuration of this material, consistent with both the collinear ferrimagnetic and canted antiferromagnetic descriptions suggested previously\cite{Tronc.1998, Gich.2005}.  Furthermore, the change in the magnetic response of the tetrahedral site at the spin reorientation temperature further supports the notion of a collective tetrahedral distortion -- in crystal structure and/or electronic localization -- to which the octahedral sites adjust.  Indeed, there may be a sufficiently strong Fe$_4$-Fe$_4$ inter-site coupling such that a spontaneous distortion to lower symmetry with cooling is energetically favoured, which is reasonable given the shared oxygen bonds in these sites. This may also result in the observed spontaneous exchange bias, since a change in crystal symmetry will surely affect the exchange interactions present, and a coexistence of two or more regions of differing exchange interactions will generally produce a unidirectional anisotropy. 

\section{Conclusions}

We find that clear evidence of the spin transition unique to $\epsilon$-Fe$_2$O$_3$, however the shift from the low temperature to high temperature magnetic configuration has been driven to lower temperatures in these $\sim$8~nm nanoparticles compared to previous reports on nanoparticles more than twice this size.
Due to finite-size effects, a significant amount of spin disorder is revealed by a spontaneous exchange bias field whose temperature dependence is linked to the transition described above, as well as the substantial frustration on ($\Gamma(T)$), and thermal evolution of the Fe sites' coordination environments ($\delta(T)$, $\Delta(T)$ and $B_{\rm hf}(T)$  (especially the Td Fe$_4$ sites), likely due to many surface sites\cite{finitesize} suffering broken coordination due to the relative smaller size and increased surface area of the 8~nm $\epsilon$-Fe$_2$O$_3$ nanoparticles.  We find that the site-specific magnetism maps onto the changes in anisotropy (tracking with the coercive field's temperature dependence) and an abrupt step in the hyperfine fields associated with the Fe$_4$ tetrahedral sites suggests strongly a change in that superexchange pathway through the O$_2$ ions is responsible. Furthermore, a change in the magnetic response of the tetrahedral site in intermediate fields at the spin reorientation temperature indicates that a collective tetrahedral distortion to which the octahedral sites adjust is occurring. The definite and unique thermal evolution of the Fe$_4$ sites along with an apparent lack of inter-site disorder suggests a collective transition of electronic localization properties, as temperature dependence x-ray diffraction measurements present no obvious change in crystal structure.  Examining the temperature dependence of the $e_g$ and $t_{2g}$ electronic states of the O ions using K-edge XAS, as well as hard x-ray K-edge XAFS investigations of the Fe coordination as a function of temperature and field would provide further insights into the structure-function relationships in this novel iron-oxide. 

\acknowledgments
We thank Dr.~D. J. Keavney for assistance with the XAS and XMCD measurements.  The authors acknowledge funding from the Natural Sciences and Engineering Research Council of Canada (RGPIN-2018-05012) and the Canada Foundation for Innovation.  Use of the Advanced Photon Source at Argonne National Laboratories was supported by the US DOE under contract DE-AC02-06CH11357. 


\bibliography{epsilonFe2O3.bib}

\begin{thebibliography}{46}
\bibitem [{fin()}]{finitesize}%
  \BibitemOpen
  \href@noop {} {}\bibinfo {note} {If we assume a 8~nm sided cube is the
  typical crystallite (in keeping with the observed TEM of these and others
  $\epsilon$-Fe$_2$O$_3$ nanoparicles) from the unit cell, we expect $\sim$70\%
  of the Fe$^3+$ ions to be on or near the surface, while for the typical 25~nm
  sized crystallies $\sim$20\% of the Fe$^{3+}$ ions are on or near the
  surface}\BibitemShut {NoStop}%
\bibitem [{XRD()}]{XRDcomment}%
  \BibitemOpen
  \href@noop {} {}\bibinfo {note} {The only observable different in XRD
  patterns was from the backgrounds from the different sample holders and
  stages.}\BibitemShut {Stop}%
\bibitem [{ffa()}]{ffactor}%
  \BibitemOpen
  \href@noop {} {}\bibinfo {note} {$f$ was calculated from spectra with
  velocity ranges $\pm$15~mm/s to avoid potential baseline
  artifcats.}\BibitemShut {Stop}%
\end{thebibliography}


\end{document}